\documentclass[12pt,amsmath,amssymb,aps,nofootinbib]{revtex4-1}
\usepackage{lipsum}
\usepackage{graphicx,epsfig,dcolumn,bm,epic,eepic,float}
\usepackage[T1]{fontenc}
\usepackage[utf8]{inputenc}
\usepackage{mathtools} 
\usepackage{pifont}
\usepackage{amsmath}
\usepackage{latexsym}
\usepackage{color}
\usepackage{cancel}
\usepackage{scrextend}
\usepackage{amsmath}
\usepackage{amsthm}
\usepackage{eucal}
\usepackage{amssymb}
\usepackage{mathrsfs}
\usepackage[bottom]{footmisc}
\usepackage{makeidx,shortvrb,latexsym}
\usepackage{epstopdf}
\usepackage{mathtools}
\usepackage{ragged2e}
\usepackage[T1]{fontenc}
\usepackage{lipsum}
\usepackage{array}
\usepackage{hyperref}
\usepackage[flushleft]{threeparttable}
\usepackage{longtable}
\usepackage{tasks}
\usepackage{blindtext}
\usepackage{enumitem}
\usepackage{multirow}

\makeatletter
\newcommand*{\rom}[1]{\expandafter\@slowromancap\romannumeral #1@}
\newcommand {\be}{\begin{equation}}
\newcommand {\ee}{\end{equation}}
\newcommand {\ba}{\begin{eqnarray}}
\newcommand {\ea}{\end{eqnarray}}
\newcommand {\bea}{\begin{eqnarray}}
\newcommand {\eea}{\end{eqnarray}}

\pdfoutput=1

\numberwithin{equation}{section}
\renewcommand{\Re}{\mathrm{Re}}
\renewcommand{\Im}{\mathrm{Im}}

\begin{document}

\title{{\LARGE Accidental Symmetries in the 2HDMEFT} \\[4mm]
${}$}

\author{\large Callum Birch-Sykes\footnote{callum.birch-sykes@manchester.ac.uk}$\,$}
\author{\large Neda Darvishi\footnote{neda.darvishi@manchester.ac.uk}$\,$}
\author{\large Yvonne Peters\footnote{yvonne.peters@manchester.ac.uk}$\,$}
\author{\large Apostolos Pilaftsis\footnote{apostolos.pilaftsis@manchester.ac.uk}$\,$}

\affiliation{~~~~~~~~~~~~~~~~~~~~~~~~~~~~~~~~~~~~~~~~~~~~~~~~~~~~~~~~~~~~${}$\vspace{-3mm}\\
Consortium for Fundamental Physics, School of Physics and Astronomy,\\
University of Manchester, Manchester M13 9PL, United
Kingdom}

\begin{abstract}
${}$

\centerline{\bf ABSTRACT} \medskip
\noindent
We construct accidentally symmetric potentials in the framework of  Two Higgs Model Effective Field Theory (2HDMEFT) including higher-order operators of dimension~6 and dimension~8. Our construction is facilitated by an earlier developed technique based on prime invariants. In addition, 
we employ an alternative method that utilises the generators of each symmetry in the bi-adjoint 
representation and show how this method can be used to identify operators of any higher dimension.  The accidentally symmetric 2HDMEFT potentials exhibit two classes of symmetries: (i) continuous symmetries and (ii) discrete symmetries. The number of continuous symmetries in the 2HDMEFT remains the same as in the 2HDM. However, the introduction of higher-order operators allows additional higher-order discrete symmetries, such as~$Z_n$ and CP$n$. We classify the full list of the 17 accidental symmetries in the 2HDMEFT including dimension-6 and dimension-8 operators, and derive the relations that govern the theoretical parameters of the corresponding effective potentials.

\end{abstract}
\maketitle
\newpage

\section{Introduction}\label{sec:intro}

The discovery of the Higgs boson at the Large Hadron Collider (LHC) \cite{HiggsAtlas,HiggsCMS} completed the last piece of the puzzle in terms of fundamental particles predicted by the Standard Model~(SM)~\cite{HiggsPredictionPHiggs,HiggsPredictionEngBrout}. Nonetheless, the SM still remains incomplete as the ultimate theory of nature, as it fails to address several key theoretical questions and cosmological observations, such as the matter-antimatter asymmetry and the origin of Dark Matter (DM) in the Universe.  Many of the theories Beyond the SM (BSM) that have been put forward in order to solve these problems utilise an extended Higgs sector. A classical example is the Minimal Supersymmetric Standard Model (MSSM) which introduces two Higgs doublets of opposite hypercharge to cancel out the gauge anomalies that would otherwise have been induced by the two higgsino doublets, namely by their fermionic supersymmetric (SUSY) partners~\cite{MSSM}. One of the simplest extensions of the SM is the Two Higgs Doublet Model (2HDM), which augments the SM scalar sector via the introduction of a second complex scalar doublet~\cite{Tviolation,PW,Ginzburg,cp-odd-nhiggs,Delgado}. This extension can provide new sources of spontaneous and explicit CP violation \cite{Tviolation,PW}, predict stable scalar DM candidates \cite{scalarphantoms,KrawczyklDM}, and give rise to electroweak baryogenesis \cite{Basymmetry, baryogenisis}. Unlike the SM, the potential of the 2HDM may exhibit many accidental symmetries, whose breaking may result in pseudo-Goldstone bosons \cite{pseudoGoldstones}, mass hierarchies, flavour-changing neutral currents \cite{neutralcurrents}, and CP violation \cite{2hdm-cpv, cpv-nhdm}. Many of these accidental symmetries can produce models with natural SM alignment, such as the Maximally Symmetric 2HDM which exhibits quartic coupling unification up to the Planck scale \cite{MS2hdm, quarticunificationMS2HDM,DPMS2HDM}.

Effective field theories (EFTs) seek to describe the low-energy dynamics of quantum field theories. This is achieved by `integrating out' heavy degrees of freedom which cannot be produced on-shell in the relevant experimental setting. This amounts to keeping only light degrees of freedom in the theory as dynamical fields~\cite{egts,eftlectures}. In the so-called Standard Model Effective Field Theory (SMEFT), the dynamical light fields are the observed particles of the SM and the integrated heavy degrees of freedom represent hypothetical new particles~\cite{Buchmuller,Grzadkowski}. The SMEFT provides a model-independent way to constrain many BSM signals, by assuming that any new particle is too heavy to be produced directly \cite{Dedes:2017zog,EFTBSM}. However, according to current bounds from collider experiments, additional Higgs bosons arising from an extended scalar sector, such as those from the 2HDM, have allowed masses near the EW scale~\cite{MSSMHiggsSearch,Hanson:2018uhf}. Therefore, it is natural to include the additional fields of the 2HDM as dynamical fields in an EFT context, as in the case of an MSSM scenario with heavy SUSY partners but light Higgs doublets~\cite{PW,Carena:2015uoe,EFT2HDM,Anisha-2HDMEFT,Banerjee}. 

 In the 2HDMEFT, the Lagrangian of the 2HDM, $\mathcal{L_\mathrm{2HDM}}$, is extended through a series of higher-order dimension-$n$ $\mathrm{SU(3)_c}\otimes\mathrm{SU(2)}_L\otimes\mathrm{U(1)}_Y$ gauge-invariant operators, $\mathcal{O}_i^{(\mathrm{dim}=n)}$, describing higher-order interactions as follows:
 \begin{equation}
     \mathcal{L}_\mathrm{2HDMEFT}=\mathcal{L}_\mathrm{2HDM}+\sum_{n>4}\sum_i\frac{C_i^{(\mathrm{dim}=n)}}{\Lambda^{n-4}}\mathcal{O}_i^{(\mathrm{dim}=n)},
 \end{equation}
where $\Lambda$ is some higher mass scale of physics beyond the 2HDM and $C_i$ stand for the corresponding dimensionless coupling constants, i.e.~the so-called Wilson coefficients. Note that operators suppressed by odd powers of $\Lambda$ usually describe baryon- and lepton-number violating terms~\cite{Buchmuller}. One drawback of using EFTs is the infinite number of parameters present. In most applications of the SMEFT, it is sufficient to include terms up to dimension~6 and ignore those suppressed by higher powers of~$\Lambda$~\cite{Appelquist}. However, for candidate 2HDM signal regions of large $\tan\beta$\footnote{$\tan\beta$ is defined as the ratio of the vacuum expectation values, $v_{1,2}$, of the Higgs doublets.}, it may be necessary to include dimension-8 operators to capture the correct analytic behaviour \cite{HiggsEFTBeyond,validitySMEFT}. 

Earlier studies have presented the full classification of a maximum of 13 accidental symmetries in the 2HDM~\cite{apilaftsis,vacuumTopology,ndarvishi}. This classification is obtained by working out all distinct subgroups of the maximal symmetry group $G^{\Phi}_\mathrm{2HDM}=\left[\mathrm{Sp(4)}/Z_2\right]\otimes\mathrm{SU(2)}_L$ or alternatively $G^R_\mathrm{2HDM}=\mathrm{SO(5)}\otimes\mathrm{SU(2)}_L$ in a bilinear field-space formalism. These accidental symmetries can be continuous or discrete in nature. Previous work has formalised a method of classifying and constructing accidentally symmetric potentials in $n$HDMs as subgroups of the maximal symmetry group $\mathrm{Sp}(2n)$, by expressing these potentials in terms of the so-called prime invariants${}$~\cite{ndarvishi}. According to this method, potentials that are invariant under the possible accidental continuous symmetries can be constructed algebraically from the fundamental building blocks respecting each symmetry.

In this paper, we extend the above formalism to algebraically construct accidentally symmetric\- effective potentials in the 2HDMEFT framework that includes higher-order operators${}$ of dimension~6 and dimension~8. The number of continuous symmetries present in the model remains the same as for the 2HDM. However, the introduction of higher-order operators allows additional higher-order discrete symmetries to be present, including new Higgs Family (HF), CP and non-Abelian symmetries. We identify all 17 accidental symmetries\- in the 2HDMEFT and pay special attention to the new symmetries, in addition to the 13 known symmetries of the 2HDM, arising from the addition of dimension-6 and dimension-8 operators to the potential. In each case, we present the relations that govern the theoretical parameters in the 2HDMEFT potential. Additionally, we recover the same relations following an alternative approach that utilises the generators of each symmetry in the bi-adjoint representation. Furhermore, we detail how this approach can be extended to operators of any higher dimension.

The layout of this paper is as follows. In Section~\ref{sec:s2}, we define the 2HDMEFT potential extended up to dimension-6 and dimension-8 operators. In Section~\ref{sec:cs}, we describe the framework for constructing accidentally symmetric effective potentials that include dimension-6 and dimension-8 operators, by employing an earlier developed method based on prime invariants${}$~\cite{ndarvishi}. In Section \ref{sec:dis}, we discuss the discrete symmetries of 2HDM and those arising from the introduction of higher-order operators to the potential. Furthermore, the combination of continuous and discrete symmetries and the full classification of the 17~accidental${}$ symmetries of 2HDMEFT and the corresponding parameter relations are given in Section~\ref{sec:full}.  Section \ref{sec:con} contains our remarks and conclusions. Technical details are delegated to Appendices \ref{ap:V}, \ref{ap:nparams} and \ref{ap:KZ} including a discussion on the free parameters in 2HDMEFT with arbitrary dimension-$2n$ operators. In addition, we present a method for computing the parameter relations among higher-order operators occurring in the effective potential for a given symmetry.  The applications of symmetries in the bilinear field-space formalism are assigned to Appendices \ref{ap:coadjoint} and~\ref{ap:dis}.

\section{The Scalar Potential in the 2HDMEFT} \label{sec:s2}
 
The 2HDM contains two scalar iso-doublets, $\phi_{1,2}$, with $\mathrm{U(1)_Y}$ hypercharges ${Y_{\phi_{1,2}}=1/2}$. In this doublet field space, the most general renormalisable 2HDM potential may conveniently  be written down as
\begin{equation}
V_\mathrm{2HDM}=\ \displaystyle\sum_{i,j=1}^{2}\, m_{ij}^2\, ( \phi_i^{\dagger} \phi_j)+ \ \displaystyle\sum_{i,j\, k,l=1}^{2}\, 
\lambda_{ijkl}\, ( \phi_i^{\dagger} \phi_j)( \phi_k^{\dagger} \phi_l),
\end{equation}
with $m^2_{ji} = m^{2\,*}_{ij}$ and $\lambda_{klij} = \lambda_{ijkl}$.  In general, the
SU(2)$_L\otimes$U(1)$_Y$ invariant 2HDM potential contains four squared mass terms, ($\mu_{1\,(2)}^2=m_{11\, (22)}^2$, $\mathrm{Re}[m_{12}^2]$ and $\mathrm{Im}[m_{12}^2]$), and ten quartic couplings, ($\lambda_{1,2,3,4}$, $\mathrm{Re}[\lambda_{5,6,7}]$ and $\mathrm{Im}[\lambda_{5,6,7}]$), where for simplicity, we used the short-hand notation $\lambda_i\, (i=1,2,\dots,10$) instead of $\lambda_{ijkl}$. In the 2HDMEFT, the 2HDM is extended through an infinite series of higher order operators. This results in an effective potential which includes higher order operators of even dimension involving the bilinear terms $\phi_i^\dagger\phi_j$. If we include terms arising from operators of dimension~6 and dimension~8 and neglect those suppressed by higher powers of $\Lambda$, we may write the 2HDMEFT effective potential as
 \begin{equation}
V_\text{2HDMEFT}=V_\mathrm{2HDM}+V^\mathrm{(dim=6)}_\mathrm{2HDM}+V^\mathrm{(dim=8)}_\mathrm{2HDM}+\mathcal{O}(\Lambda^{-6}),
 \end{equation}
 with
 \begin{align}
V^\mathrm{(dim=6)}_\mathrm{2HDM}=
 \frac{1}{\Lambda^2}\ \displaystyle\sum_{i,j\, k,l\, m,n=1}^{2}\, 
\kappa_{ijklmn}\, ( \phi_i^{\dagger} \phi_j)( \phi_k^{\dagger} \phi_l)( \phi_m^{\dagger} \phi_n),
\end{align}
and
\begin{align}
    V^\mathrm{(dim=8)}_\mathrm{2HDM}=\frac{1}{\Lambda^4}\ \displaystyle\sum_{i,j\, k,l\, m,n,p,q=1}^{2}\, 
\zeta_{ijklmnpq}\, ( \phi_i^{\dagger} \phi_j)( \phi_k^{\dagger} \phi_l)(\phi_m^{\dagger} \phi_n)(\phi_p^{\dagger} \phi_q).
\end{align}
This introduces additional 20 hexic couplings, ($\kappa_{1,2,\dots,6}$, $\mathrm{Re}\big[\kappa_{7,8,\dots,13}\big]$ and  $\mathrm{Im}\big[\kappa_{7,8,\dots,13}\big]$), and 35 octic couplings ($\zeta_{1,2,\dots,9}$, $\mathrm{Re}\left[\zeta_{10,11,\dots,22}\right]$ and $\mathrm{Im}\left[\zeta_{10,11,\dots,22}\right]$), using the short-hand notation $\kappa_i\, (i=1,2,\dots, 13$) and $\zeta_i\, (i=1,2,\dots, 22$) instead of $\kappa_{ijklmn}$ and $\zeta_{ijklmnpq}$, respectively. Thereby,  the effective potential now contains 69 parameters. The explicit expression for the 2HDMEFT potential and a discussion of general higher-order dimension-$2n$ operators are given in Appendices~\ref{ap:V} and~\ref{ap:nparams}, respectively. 
 
In an equivalent way, the 2HDMEFT potential can also be expressed in a bilinear field-space formalism~\cite{stability-symv-2hdm,minkowskipotential,potential-nhdm,minkowski2,
vacuumTopology}. To establish this, we first define a complex 8-dimensional (8D) Majorana multiplet, $\mathbf{\Phi}$, as 
\begin{equation}
\mathbf{\Phi}=\begin{pmatrix}\phi_1\\\phi_2\\i\sigma^2\phi_1^*\\i\sigma^2\phi_2^*\end{pmatrix},
\end{equation}
where $i\sigma^2\phi^*_j$ is the hypercharge conjugate of $\phi_j$. The $\mathbf{\Phi}$-multiplet transforms covariantly under $\mathrm{SU(2)}_L$ gauge transformations as, $\mathbf{\Phi}\to\mathrm{U}_L\mathbf{\Phi}$ with $\mathrm{U}_L\in\mathrm{SU(2)}_L$, and also satisfies the Majorana property, $\mathbf{\Phi}=C\mathbf{\Phi}^*$ with $C=\sigma^2\otimes\sigma^0\otimes\sigma^2$.
 
 Using the $\mathbf{\Phi}$-multiplet, we may define a bilinear field 6-vector,
\begin{equation}
    \label{eq:RAvector}
    R^A\equiv \mathbf{\Phi}^{\dagger}\Sigma^A\mathbf{\Phi}=\begin{pmatrix}\phi^{\dagger}_1\phi_1+\phi^{\dagger}_2\phi_2\\\phi^{\dagger}_1\phi_2+\phi^{\dagger}_2\phi_1\\-i\Big[\phi^{\dagger}_1\phi_2-\phi^{\dagger}_2\phi_1\Big]\\\phi^{\dagger}_1\phi_1-\phi^{\dagger}_2\phi_2\\\phi_1^\top i\sigma^2\phi_2-\phi_2^{\dagger}i\sigma^2\phi_1^*\\-i\Big[\phi_1^\top i\sigma^2\phi_2+\phi_2^{\dagger}i\sigma^2\phi_1^*\Big]\end{pmatrix},
\end{equation}
with $A=0,1,..,5$. The 6-vector, $R^A$, is $\mathrm{SU(2)}_L$ invariant and forms a null $\mathrm{SO(1,5)}$ vector. The matrices ${\Sigma}^{A}$ have $8 \times 8$ elements and
can be expressed in terms of double tensor products as,
\begin{align}\nonumber
    \Sigma^{0,1,3}&=\frac{1}{2}\sigma^0\otimes\sigma^{0,1,3}\otimes\sigma^0,  &&\Sigma^2=\frac{1}{2}\sigma^3\otimes\sigma^2\otimes\sigma^0, \\  \Sigma^{4}&=-\frac{1}{2}\sigma^{2}\otimes\sigma^2\otimes\sigma^0,
    &&\Sigma^{5}=-\frac{1}{2}\sigma^{1}\otimes\sigma^2\otimes\sigma^0.
\end{align}

In the $R^A$-space defined in~\eqref{eq:RAvector}, the renormalisable 2HDM potential takes on the quadratic form \cite{apilaftsis} \begin{equation}
  V_\mathrm{2HDM}=-\frac{1}{2}M_AR^A+\frac{1}{4}L_{AB}R^AR^B,
\end{equation}
where $M_A$ is a six-dimensional mass vector and $L_{AB}$ is a $6\times6$ dimensional quartic coupling matrix. The correspondence between the elements of $M_A$ and $L_{AB}$ in the 
bilinear field space, and 
the theoretical parameters in the linear field space is
\begin{equation}
    M_A=(\mu^2_1+\mu^2_2,\ 2\,\mathrm{Re}[m^2_{12}],\ -2\,\mathrm{Im}[m_{12}^2],\ \mu_1^2-\mu^2_2,\ 0,\ 0),
\end{equation}
and
\begin{equation}
L_{AB}=\begin{pmatrix}\lambda_1+\lambda_2+\lambda_3&\mathrm{Re}[\lambda_6+\lambda_7]&-\mathrm{Im}[\lambda_6+\lambda_7]&\lambda_1-\lambda_2&0&0\\\mathrm{Re}[\lambda_6+\lambda_7]&\lambda_4+\mathrm{Re}[\lambda_5]&-\mathrm{Im}[\lambda_5]&\mathrm{Re}[\lambda_6-\lambda_7]&0&0\\-\mathrm{Im}[\lambda_6+\lambda_7]&-\mathrm{Im}[\lambda_5]&\lambda_4-\mathrm{Re}[\lambda_5]&-\mathrm{Im}[\lambda_6-\lambda_7]&0&0\\\lambda_1-\lambda_2&\mathrm{Re}[\lambda_6-\lambda_7]&-\mathrm{Im}[\lambda_6-\lambda_7]&\lambda_1+\lambda_2-\lambda_3&0&0\\0&0&0&0&0&0\\0&0&0&0&0&0\end{pmatrix},
\end{equation}
where the elements with $A,B\in\{4,5\}$ vanish as they violate $\mathrm{U(1)}_Y$. In this bilinear representation, we can write the additional higher order terms of the 2HDMEFT effective potential in the following form:
 \begin{align}
     V_\mathrm{2HDM}^\mathrm{(dim=6)}=&\ \frac{1}{\Lambda^2}K_{ABC}R^AR^BR^C,\\[3mm]
     V^\mathrm{(dim=8)}_\mathrm{2HDM}=&\ \frac{1}{\Lambda^4}Z_{ABCD}R^AR^BR^CR^D,
 \end{align}
 where $K_{ABC}$ and $Z_{ABCD}$ are symmetric rank-3 and rank-4 tensors describing the hexic and octic couplings, respectively. The relations of the elements of $K_{ABC}$ and $Z_{ABCD}$ with the couplings in the original linear field-space are given in Appendix~\ref{ap:KZ}.

 The gauge kinetic term of the 2HDM can be written in terms of the $\mathbf{\Phi}$-multiplet as
 \begin{equation}
     T=\frac{1}{2}(D_\mu\mathbf{\Phi})^\dagger(D^\mu\mathbf{\Phi}),
 \end{equation}
 where the covariant derivative, $D_\mu$, in the $\mathbf{\Phi}$-space  is given by
 \begin{equation}
D_{\mu} = \sigma^{0} \otimes \sigma^{0} \otimes \bigg(\sigma^0 \partial_\mu +i g_w W^i_\mu{\sigma^i\over 2}\,\bigg)+\sigma^3\otimes \sigma^0\otimes i\frac{g_Y}{2}B_\mu \sigma^0.
 \end{equation}
 The maximal symmetry acting on $T$ that leaves the $\mathrm{SU(2)}_L$ gauge kinetic term invariant is
 \begin{equation}
     G_\mathrm{2HDM}^\mathbf{\Phi}=\left[\mathrm{Sp(4)}/Z_2\right]\otimes\mathrm{SU(2)}_L,
 \end{equation}
which means that the local SU(2$)_L$ group generators commute with all generators of~Sp(4).  Note that the maximal symmetry is an approximate symmetry of $T$ which becomes an exact maximal symmetry in the limit $g_Y\to\,0$. In the case of 2HDMEFT potentials, this symmetry is also the maximal symmetry, as any additional gauge kinetic terms can be set to zero through symmetry arguments.  The generators of the global symmetry can be defined either in the $\mathbf{\Phi}$-space or equivalently in the $R^A$-space.  In the latter case, the $R^A$-vector transforms in the bi-adjoint representation of the bilinear formalism~\cite{ndarvishi}. The applications of symmetries in the bilinear basis including the adjoint and bi-adjoint representations of the Sp(4) group are discussed in Appendix~\ref{ap:coadjoint}.

 Knowing that Sp(4) is the maximal symmetry group allows us to classify all SU(2)$_L$-preserving accidental symmetries of the 2HDMEFT potentials. The symmetric potentials can be constructed in the original linear field space by means of bilinears which manifestly respect two distinct classes of symmetries: (i) continuous symmetries and (ii) discrete symmetries${}$. These will be discussed in the following sections.

\section{Continuous Symmetries and Prime-Invariants in 2HDMEFT}\label{sec:cs}
 
Previous work has formalised an approach to algebraically classifying and constructing accidentally symmetric $n$HDM potentials by virtue of the so-called prime invariants \cite{ndarvishi}. Prime invariants allow us to systematically construct symmetric potentials from fundamental building blocks which respect those symmetries. To define prime invariants in the 2HDMEFT, we must first consider the possible continuous symmetries that are contained as maximal subgroups of $\mathrm{Sp(4)}$. In this way, the following 7 possible $\mathrm{U(1)}_Y$-conserving continuous accidental symmetries are obtained:
 \begin{align}
    \label{eq:SP4subgroups}
 &(a)\quad\mathrm{Sp}(4)                         & (e)&\quad\mathrm{U(1)_{PQ}}\otimes\mathrm{Sp}(2)_{}\nonumber\\
 &(b)\quad\mathrm{Sp}(2)\otimes\mathrm{Sp}(2)     & (f)&\quad\mathrm{U(1)_{PQ}}\otimes\mathrm{U(1)}_{Y} \nonumber\\
 &(c)\quad\mathrm{Sp}(2)                         &(g)&\quad\mathrm{SO(2)_{HF}}\otimes\mathrm{U(1)}_{Y} \nonumber\\
 &(d)\quad\mathrm{SU(2)_{HF}}\otimes\mathrm{U(1)}_{Y}  
 \end{align}

The quantities that are invariant under the various groups listed in~\eqref{eq:SP4subgroups} form a set of prime invariants${}$. In the 2HDMEFT, the prime invariants pertinent to building accidentally symmetric effective potentials are:
\begin{itemize}

\item[(a)] The maximal prime invariant, $S$, which is constructed from the Majorana multiplet as
\begin{equation}
    S=\mathbf{\Phi}^\dagger\mathbf{\Phi}.
\end{equation}
Note that $S$ is invariant under the maximal symmetry group Sp(4) and all of its maximal subgroups. 
In addition to the above maximal prime invariant, there are minimal prime invariants which are invariant under $\mathrm{Sp(2)}\subset\mathrm{Sp(4)}$, i.e.
 \begin{align} S_{12} =&\ \begin{pmatrix}\phi_1\\i\sigma^2\phi_1^*\end{pmatrix}^\dagger\begin{pmatrix}\phi_2\\i\sigma^2\phi_2^*\end{pmatrix}\ =\ \phi_1^\dagger\phi_2+\phi_2^\dagger\phi_1\,,\\[3mm]   
S_{11} =&\ \begin{pmatrix}\phi_1\\i\sigma^2\phi_1^*\end{pmatrix}^\dagger\begin{pmatrix}\phi_1\\i\sigma^2\phi_1^*\end{pmatrix}\ =\ \phi_1^\dagger\phi_1\,.
\end{align}

\item[(b)] Potential terms that are invariant  under $\mathrm{SU(2)}_{\rm HF}\otimes\mathrm{U(1)}_Y$ 
and $\mathrm{Sp(2)}_{\phi_1 \phi_2}$ in the bases $(\phi_i\,,\, \phi_j)$ and 
$(\phi_i\,,\, i \sigma^2 \phi_j^*)$ may be expressed in terms of the quantities $D^2_{12}=D^a_{12}D^a_{12}$ and $D'^2_{12}=D'^a_{12}D'^a_{12}$, respectively, where
 \begin{align}
D^a_{12}&=\begin{pmatrix}\phi_1\\\phi_2\end{pmatrix}^\dagger\sigma^a\begin{pmatrix}\phi_1\\\phi_2\end{pmatrix}\, =\: \phi_1^\dagger\sigma^a\phi_1+\phi_2^\dagger\sigma^a\phi_2\,, \\[2mm]
D'^a_{12}&=\begin{pmatrix}\phi_1\\i\sigma^2\phi_2^*\end{pmatrix}^\dagger\sigma^a
\begin{pmatrix}\phi_1\\i\sigma^2\phi_2^*\end{pmatrix}\, =\:  \phi_1^\dagger\sigma^a\phi_1-\phi_2^\dagger\sigma^a\phi_2\,.
\end{align}

\item[(c)] Finally, $\mathrm{SO(2)}$-invariant potential terms may be constructed 
using the quantity $T^2_{12}=T_{12}T_{12}^*$, with
\begin{equation}
    T_{12}=\phi_1\phi_1^\top+\phi_2\phi_2^\top.
\end{equation}
\end{itemize}

Having defined the prime invariants, we can now show how they may be used to construct accidentally symmetric potentials in the framework of 2HDMEFT. The renormalisable part of the potential built from prime invariants is
 \begin{align}
     V_{\rm sym}\: =\: -\mu^2S+\lambda_SS^2+\lambda_DD^2+\lambda_TT^2,
 \end{align}
where each term is invariant under the given symmetry (a), (b) or (c) mentioned above. This approach can be extended to construct higher dimensional terms of symmetric 2HDMEFT potentials, e.g.
 \begin{align}
     V_{\rm sym}^\mathrm{(dim=6)}=&\ \frac{1}{\Lambda^2}(\kappa_SS^3+\kappa_{D}D^2S+\kappa_{T}T^2S)\,,\\[2mm]
V^\mathrm{(dim=8)}_{\rm sym}=& \ \frac{1}{\Lambda^4}(\zeta_SS^4+\zeta_{D}D^4+\zeta_TT^4+\zeta_{SD}S^2D^2+\zeta_{ST}S^2T^2
+\zeta_{DT}D^2T^2)\,. 
\end{align}
Hence, the following accidentally symmetric potentials can be constructed\footnote{ \noindent The subscript $\phi_i$ indicates a transformation acting only on $\phi_i$, the subscript $\phi_1\phi_2$ indicates a transformation between $\phi_1$ and $\phi_2$ and the subscript $\phi_1+\phi_2$ indicates a transformation on both $\phi_1$ and $\phi_2$.}:
\begin{tasks}(2)
\task   $\mathrm{Sp(4)}$: $V[S_{11}+S_{22}]$
\task     $\mathrm{SU(2)}$: $V[S_{11}+S_{22}, D_{12}^2]$
\task    $\mathrm{Sp(2)}_{\phi_1} \otimes \mathrm{Sp(2)}_{\phi_2}$: $V[S_{11},S_{22}]$
\task     $\mathrm{Sp(2)}_{\phi_1 +\phi_2}$: $V[S_{11}, S_{22}, S_{12} ]$
\task    $\mathrm{U(1)} \otimes \mathrm{Sp(2)}_{\phi_1 \phi_2}$: $V[S_{11}+S_{22}, D_{12}^{\prime 2}]$
\task    $\mathrm{CP1}\otimes\mathrm{SO(2)}$: $V[S_{11}+S_{22}, D_{12}^2, T^2_{12}]$
\end{tasks}
There is another continuous symmetry present in the 2HDMEFT, the Abelian Peccei--Quinn~(PQ) symmetry, $\mathrm{U(1)_{PQ}}$~\cite{PQ}.  A~PQ-symmetric potential is invariant under the transformations, $\phi_1\to e^{i\alpha}\phi_1$ and $\phi_2\to e^{-i\alpha}\phi_2$. This can be constructed simply by considering\- only such combinations of $\phi_i^\dagger\phi_j$ that cancel out this phase.

We obtain all possible symmetric effective potentials, extended up to dimension-6 and dimension-8 operators, under the accidental continuous symmetries of the 2HDMEFT identified\- above.  The corresponding parameter relations for each potential are shown in Table~\ref{tab:cont}. Note that there are no new continuous symmetries in 2HDMEFT with respect to 
the dimension-4 2HDM potential.

\newcolumntype{t}{>{\tiny}l}
\newcolumntype{s}{>{\scriptsize}c}
\addtocounter{table}{-1}
\begin{table}[t]
\begin{longtable}[h]{ | s | t |}
\hline
\small{Symmetry} &  \small{Non-zero parameters of Symmetric 2HDMEFT Potential}\\
\hline \hline

$\mathrm{U(1)_{PQ}}$ & \begin{tabular}{@{}l@{}} $\mu_1^2$, $\mu_2^2$, $\lambda_1$, $\lambda_2$, $\lambda_3$, $\lambda_4$\\ $\kappa_1$, $\kappa_2$, $\kappa_3$, $\kappa_4$, $\kappa_5$, $\kappa_6$\\ $\zeta_1$, $\zeta_2$, $\zeta_3$, $\zeta_4$, $\zeta_5$, $\zeta_6$, $\zeta_7$, $\zeta_8$, $\zeta_9$ \end{tabular} \\ \hline

$\mathrm{CP1}\otimes\mathrm{SO(2)_{\rm HF}}$ &  \begin{tabular}{@{}l@{}}$\mu_1^2=\mu_2^2$, $\lambda_1=\lambda_2$, $\lambda_3$, $\lambda_4$, Re($\lambda_5)=2\lambda_1-\lambda_{34}$\\ $\kappa_1=\kappa_2$, $\kappa_3=\kappa_4$, $\kappa_5=\kappa_6$, Re$(\kappa_8)=$Re$(\kappa_9)=\frac{1}{2}(3\kappa_1-\kappa_3-\kappa_5)$ \\$\zeta_1=\zeta_2$, $\zeta_3$, $\zeta_4=\zeta_5$, $\zeta_6$, $\zeta_7=\zeta_8$, $\zeta_9$, \\
$\mathrm{Re}(\zeta_{10})=-\frac{1}{4}\Re(\zeta_{13})+\frac{1}{2}\Re(\zeta_{14})-\frac{1}{4}\Re(\zeta_{16})$,
$\mathrm{Re}(\zeta_{13})=\frac{1}{6}(4\zeta_1+2\zeta_3-4\zeta_4-4\zeta_6+2\zeta_7-\zeta_9)$ \\$\mathrm{Re}(\zeta_{14})=\mathrm{Re}(\zeta_{15})=\frac{1}{2}(4\zeta_1-\zeta_4-\zeta_7)$,
$\mathrm{Re}(\zeta_{16})=\frac{1}{2}(4\zeta_1-2\zeta_3+2\zeta_4-\zeta_9)$
\end{tabular}\\ \hline

$\mathrm{SU(2)_{HF}}$ & \begin{tabular}{@{}l@{}} $\mu_1^2=\mu_2^2$, $\lambda_1=\lambda_2$, $\lambda_3$, $\lambda_4=2\lambda_1-\lambda_3$ \\$\kappa_1=\kappa_2$, $\kappa_3=\kappa_4$, $\kappa_5=\kappa_6=3\kappa_1-\kappa_3$ \\ $\zeta_1=\zeta_2$, $\zeta_3$, $\zeta_4=\zeta_5$, $\zeta_6=2\zeta_1+\zeta_3-2\zeta_4$, $\zeta_7=\zeta_8=4\zeta_1-\zeta_4$, $\zeta_9=4\zeta_1-2\zeta_3+2\zeta_4$  \end{tabular}\\ \hline \hline

$\mathrm{Sp(2)}_{\phi_1+\phi_2}$ & \begin{tabular}{@{}l@{}}  $\mu_1^2$, $\mu_2^2$, Re($m_{12}^2$), $\lambda_1$, $\lambda_2$, $\lambda_3$, $\lambda_4=\text{Re}(\lambda_5)$, Re($\lambda_6$), Re($\lambda_7$) \\ $\kappa_1$, $\kappa_2$, $\kappa_3$, $\kappa_4$, $\kappa_5=2\mathrm{Re}(\kappa_8)$, $\kappa_6=2\mathrm{Re}(\kappa_9)$, $\mathrm{Re}(\kappa_7)=\frac{1}{3}\mathrm{Re}(\kappa_{10})$, Re($\kappa_{11}$), Re($\kappa_{12}$), Re($\kappa_{13}$) \\ 
$\zeta_1$, $\zeta_2$, $\zeta_3$, $\zeta_4$, $\zeta_5$, $\zeta_6=6\mathrm{Re}(\zeta_{10})=\frac{3}{2}\mathrm{Re}(\zeta_{13})$,
$\zeta_7=2\Re(\zeta_{14})$, $\zeta_8=2\Re(\zeta_{15})$, $\zeta_9=2\Re(\zeta_{16})$,  \\ $\Re(\zeta_{11})=\frac{1}{3}\Re(\zeta_{17})$, $\Re(\zeta_{12})=\frac{1}{3}\Re(\zeta_{18})$, $\Re(\zeta_{19})$, $\Re(\zeta_{20})$, $\Re(\zeta_{21})$, $\Re(\zeta_{22})$    \end{tabular} \\ \hline

$\mathrm{U(1)_{PQ}}\otimes\mathrm{Sp(2)}_{\phi_1\phi_2}$ & \begin{tabular}{@{}l@{}}  $\mu_1^2=\mu_2^2$, $\lambda_1=\lambda_2=\frac{1}{2}\lambda_3$, $\lambda_4$\\ $\kappa_1=\kappa_2=\frac{1}{3}\kappa_3=\frac{1}{3}\kappa_4$, $\kappa_5=\kappa_6$ \\$\zeta_1=\zeta_2=\frac{1}{6}\zeta_3=\frac{1}{4}\zeta_4=\frac{1}{4}\zeta_5$, $\zeta_6$, $\zeta_7=\zeta_8=\frac{1}{2}\zeta_9$ \end{tabular}\\ \hline

$\mathrm{Sp(2)}_{\phi_1}\otimes\mathrm{Sp(2)}_{\phi_2}$ & \begin{tabular}{@{}l@{}}  $\mu_1^2$, $\mu_2^2$, $\lambda_1$, $\lambda_2$, $\lambda_3$ \\
$\kappa_1$, $\kappa_2$, $\kappa_3$, $\kappa_4$ \\
$\zeta_1$, $\zeta_2$, $\zeta_3$, $\zeta_4$, $\zeta_5$     \end{tabular}\\ \hline

$\mathrm{Sp(4)}$ & \begin{tabular}{@{}l@{}}  $\mu_1^2=\mu_2^2$, $\lambda_1=\lambda_2=\frac{1}{2}\lambda_3$ \\$\kappa_1=\kappa_2=\frac{1}{3}\kappa_3=\frac{1}{3}\kappa_4$,\\ $\zeta_1=\zeta_2=\frac{1}{6}\zeta_3=\frac{1}{4}\zeta_4=\frac{1}{4}\zeta_5$ \end{tabular}\\ \hline

\end{longtable}
\caption{\it Parameter relations for the symmetric 2HDMEFT potentials, including dimension-6 and dimension-8 operators, under the possible continuous accidental symmetries. Note that the last four entries which involve symplectic groups are custodially symmetric.}
\label{tab:cont}
\end{table}
\addtocounter{table}{-1}

\section{Discrete Symmetries in 2HDMEFT}\label{sec:dis}

There are several types of discrete symmetries which can be imposed on the potential as subgroups of continuous symmetries. Common examples of symmetries of this type are the Cyclic group $Z_n$, the Permutation group $S_n$, the standard CP symmetry or possible combinations of them. In this section, we will discuss the possible discrete symmetries present in the 2HDMEFT potential up to dimension-6 and dimension-8 operators.

Let us first turn our attention to HF transformations which act on the HF field space, $(\phi_1,\; \phi_2)$. Let us consider the Abelian discrete symmetry group, $Z_n = \{1,\omega_n,\dots,(\omega_n)^{n-1}\}$, with $(\omega_n)^n=1$. Under $Z_n$, 
the Higgs doublets transform as
  \begin{align}
     Z_n:& \qquad \phi_1\to\phi_1\,, \qquad \phi_2\to \omega_n \phi_2\,,
   \end{align}
with $ \omega_n =e^{2\pi i/n}$. In the 2HDM, the $Z_2$ symmetry is usually imposed to ensure the absence of flavour changing neutral currents. In the 2HDMEFT, the addition of higher order operators allows for the presence of higher order discrete symmetries. In general, the addition of dimension-$2n$ terms to the effective potential gives rise to the presence of $Z_n$ symmetries. Hence, by including dimension-6 and dimension-8 operators, the $Z_3$ and $Z_4$ symmetries become distinct symmetries of the effective potential. In the HF field space, the $Z_n$ symmetry is 
represented by
\begin{align}
    \delta_{Z_n}=\begin{pmatrix} 1 & 0 \\ 0 & \omega_n \end{pmatrix}.
\end{align}
As a consequence, in the $\mathbf{\Phi}$-space, the generators of $Z_n$ are given by
\begin{align}
    \Delta_{Z_n}=\begin{pmatrix} \delta_{Z_n} & \mathbf{0} \\ \mathbf{0}  & \delta_{Z_n}^* \end{pmatrix}\otimes\sigma^0.
\end{align}

The Permutation group, $S_2$, is a group isomorphic to $Z_2$, so transformations arising from~$S_2$ are related to $Z_2$ transformations via a change of basis: $\phi_1'=(\phi_1+\phi_2)/\sqrt{2}$, $\phi_2'=(\phi_1-\phi_2)/\sqrt{2}$. However, the $S_2$ group can be utilised via a semi-direct product with $Z_n$ to form the non-Abelian Tetrahedral symmetry group $D_n$, such as 
\begin{equation}
    D_n\: =\: Z_n\rtimes S_2.
\end{equation}
The Tetrahedral group generators in the HF field space are given by the set,
\begin{equation}
  \delta_{S_2}=\sigma^1=\begin{pmatrix} 0 & 1 \\ 1 & 0 \end{pmatrix}\,, \qquad \delta_{Z_n}\,, \qquad  \delta_{Z_n}\delta_{S_2}.
\end{equation}
 In the $\mathbf{\Phi}$-space, the generators of these discrete symmetries may be represented
by double tensor products, i.e.
\begin{equation}
    \Delta_{S_2}=\sigma^0\otimes\sigma^1\otimes\sigma^0\,, \qquad \Delta_{Z_n}\,, \qquad  \Delta_{Z_n}\Delta_{S_2}.
\end{equation}

In general, the presence of dimension-$2n$ operators in the 2HDMEFT potential allows the presence of $D_n$ symmetries. Hence, by going to dimension~6 and dimension~8, we are able, in principle, to define $D_3$ and $D_4$ symmetric potentials.  However, there are some caveats here, as some symmetries may not yield unique potentials or may be related to another symmetric potential via a change of basis. This will be discussed further in Section V. The parameter relations corresponding to 2HDMEFT potentials which are invariant under the discrete HF symmetries, with dimension-6 and dimension-8 operators, are given in Table \ref{tab:HF}.

\begin{table}[t]

\begin{longtable*}[h]{ | s | t | s |}
\hline
 \small{Symmetry} &  \small{Non-zero parameters of Symmetric 2HDMEFT Potential} & \small{Dim} \\
\hline \hline

 $Z_2$ &  \begin{tabular}{@{}l@{}} $\mu_1^2$, $\mu_2^2$, $\lambda_1$, $\lambda_2$, $\lambda_3$, $\lambda_4$, $\lambda_5$    \\ $\kappa_1$, $\kappa_2$, $\kappa_3$, $\kappa_4$, $\kappa_5$, $\kappa_6$, $\kappa_8$, $\kappa_9$ \\$\zeta_1$, $\zeta_2$, $\zeta_3$, $\zeta_4$, $\zeta_5$ $\zeta_6$, $\zeta_7$, $\zeta_8$, $\zeta_9$, $\zeta_{10}$, $\zeta_{13}$, $\zeta_{14}$, $\zeta_{15}$, $\zeta_{16}$ \end{tabular} & $D\geq4$  \\  \hline
 
 $Z_2\rtimes S_2$ &  \begin{tabular}{@{}l@{}} $\mu_1^2=$, $\mu_2^2$, $\lambda_1=\lambda_2$, $\lambda_3$, $\lambda_4$, Re($\lambda_5$)    \\ $\kappa_1=\kappa_2$, $\kappa_3=\kappa_4$, $\kappa_5=\kappa_6$, $\kappa_8=\kappa_9^*$ \\$\zeta_1=\zeta_2$, $\zeta_3$, $\zeta_4=\zeta_5$ $\zeta_6$, $\zeta_7=\zeta_8$, $\zeta_9$, Re($\zeta_{10}$), Re($\zeta_{13}$), $\zeta_{14}=\zeta_{15}^*$, Re($\zeta_{16})$ \end{tabular} & $D\geq4$  \\  \hline

 $Z_3$ & \begin{tabular}{@{}l@{}}  $\mu_1^2$, $\mu_2^2$, $\lambda_1$, $\lambda_2$, $\lambda_3$, $\lambda_4$  \\ $\kappa_1$, $\kappa_2$, $\kappa_3$, $\kappa_4$, $\kappa_5$, $\kappa_6$, $\kappa_7$ \\$\zeta_1$, $\zeta_2$, $\zeta_3$, $\zeta_4$, $\zeta_5$, $\zeta_6$, $\zeta_7$, $\zeta_8$, $\zeta_9$, $\zeta_{11}$, $\zeta_{12}$ \end{tabular} &  $D\geq6$   \\  \hline

 $D_3$ & \begin{tabular}{@{}l@{}}   $\mu_1^2=\mu_2^2$, $\lambda_1=\lambda_2$, $\lambda_3$, $\lambda_4$  \\ $\kappa_1=\kappa_2$, $\kappa_3=\kappa_4$, $\kappa_5=\kappa_6$, Re($\kappa_7$) \\ $\zeta_1=\zeta_2$, $\zeta_3$, $\zeta_4=\zeta_5$, $\zeta_6$, $\zeta_7=\zeta_8$, $\zeta_9$, $\zeta_{11}=\zeta_{12}^*$  \end{tabular} & $D\geq6$ \\ \hline
 
 $Z_4$ & \begin{tabular}{@{}l@{}}  $\mu_1^2$, $\mu_2^2$, $\lambda_1$, $\lambda_2$, $\lambda_3$, $\lambda_4$ \\ $\kappa_1$, $\kappa_2$, $\kappa_3$, $\kappa_4$, $\kappa_5$, $\kappa_6$ \\ $\zeta_1$, $\zeta_2$, $\zeta_3$, $\zeta_4$, $\zeta_5$, $\zeta_6$, $\zeta_7$, $\zeta_8$, $\zeta_9$, $\zeta_{10}$ \end{tabular} & $D\geq8$  \\  \hline

 $D_4$ & \begin{tabular}{@{}l@{}}   $\mu_1^2=\mu_2^2$, $\lambda_1=\lambda_2$, $\lambda_3$, $\lambda_4$ \\$\kappa_1=\kappa_2$, $\kappa_3=\kappa_4$, $\kappa_5=\kappa_6$ \\$\zeta_1=\zeta_2$, $\zeta_3$, $\zeta_4=\zeta_5$, $\zeta_6$, $\zeta_7=\zeta_8$, $\zeta_9$, $\Re(\zeta_{10})$   \end{tabular} & $D\geq8$  \\ \hline

\end{longtable*}

\caption{\it Parameter relations for the symmetric 2HDMEFT potentials, including dimension-6 and dimension-8 operators, under the possible discrete HF symmetries. The 3rd column indicates what higher dimension operators must be included for the symmetry to be distinct.}
\label{tab:HF}
\end{table}
\addtocounter{table}{-1}

In addition to above HF symmetries, a class of discrete symmetries that relates ${{\phi}}_i\,\to\, {{\phi}}_{j}^*$ 
is the Generalized CP (GCP) transformations, defined as
\begin{equation}
\text{GCP}[{{\phi}}_i]=G_{ij} {{\phi}}_{j}^*,
\end{equation}
with $\{G_{ij}\} \in \text{SU(}n)\otimes$U(1). The GCP transformations realize different types of CP symmetry. In the case of the 2HDM, there are 
two types of CP symmetries: (i)~standard CP or CP1 and 
(ii)~non-standard CP or CP2, whose transformations are defined as~\cite{GCP}
\begin{align}
    \mathrm{CP1}:& \qquad \phi_i\to\phi_i^*,\\
    \mathrm{CP2}:& \qquad \phi_1\to\phi_2^*, \quad \phi_2\to-\phi_1^*.
 \end{align}
In the $\mathbf{\Phi}$-space, the generators of these discrete symmetries 
may be  given by the following set:
\begin{equation}
\Delta_{\text{CP1}}=\sigma^2 \otimes \sigma^0 \otimes \sigma^2\,, \qquad
\Delta_{\text{CP2}}=\Delta_\text{CP1}\Delta_{S_2}\Delta_{Z_2}\,.
\end{equation}

Let us now consider operators of higher dimensionality in the 2HDMEFT potential. Interestingly${}$ enough, the presence of higher order operators allows the realisation of higher discrete CP symmetries. To this end, we define the CP$n$ transformation
\begin{equation}
    \text{CP}n: \qquad  \phi_1\to \phi_2^*, \quad \phi_2\to \omega_n\phi_1^*,
\end{equation}
with $\omega_n=e^{2\pi i/n}$.
In the $\mathbf{\Phi}$-space, CP$n$ transformations may be generated by the product of the generators of CP1, $S_2$ and $Z_n$,
\begin{equation}
    \Delta_{\text{CP}n}\: =\: \Delta_\text{CP1}\Delta_{S_2}\Delta_{Z_n}\,.
\end{equation}

In general, the addition of dimension-$2n$ terms to the 2HDMEFT potential gives rise to the occurence of CP$n$ symmetries. Hence, by including dimension-6 and dimension-8 operators, we may be able to realise the CP3, CP4 and CP6 symmetries\footnote{Not to be confused with the CP3 transformation defined as $\text{CP1}\otimes\mathrm{SO(2)_{HF}}$ 
in some literature.}. Since CP6 is isomorphic to $\text{CP2}\otimes Z_3$, we do not consider this as a new CP symmetry.  The parameter relations corresponding to potentials which are invariant under the discrete CP symmetries present in 2HDMEFT, with dimension-6 and dimension-8 operators, are given in Table \ref{tab:CP}. Note that the symmetries $\text{CP2}\otimes Z_n$ are related to the $\text{CP}n$ symmetries via a reduction of the basis obtained by a $\mathrm{U(2)}$ reparameterisation  of the doublets.  Thus, additional care must be taken to identify which symmetries should be regarded as new with respect to already classified symmetries that could be obtained through a change of basis.
 
It is worth noting that the generators of CP$n$ symmetries are of order $2n$, i.e.~$\Delta_{\text{CP}n}^{2n}= {\bf 1}_8$. However, in the bilinear $R^{A}$-space, the order of both transformation matrices of CP1 and CP2 is 2, i.e.~$D^2_{\text{CP}1}= {\bf 1}_5$ and 
$D^2_{\text{CP}2}= {\bf 1}_5$.  On the other hand, the order of CP3 and CP4 transformation matrices are  $D^6_{\text{CP}3} = {\bf 1}_5$ and  $D^4_{\text{CP}4} = {\bf 1}_5$. The transformation matrices of the discrete symmetries in the bilinear $R^{A}$-space are given in Appendix~\ref{ap:dis}.

\begin{table}[t]

\begin{longtable*}[h]{ | s | t | s |}
\hline
 \small{Symmetry} &  \small{Non-zero parameters of Symmetric 2HDMEFT Potential} &\small{Dim} \\
\hline \hline

 $\mathrm{CP1}$ & \begin{tabular}{@{}l@{}}  $\mu_1^2$, $\mu_2^2$, Re($m_{12}^2$), $\lambda_1$, $\lambda_2$, $\lambda_3$, $\lambda_4$, Re($\lambda_5$), Re($\lambda_6$), Re($\lambda_7$)  \\  $\kappa_1,\dots,\,\kappa_6$, Re($\kappa_7,\dots,\,\kappa_{13}$) \\ $\zeta_1,\dots,\,\zeta_9$, Re($\zeta_{10},\dots,\,\zeta_{22}$)
 \end{tabular} &$D\geq4$\\ \hline

  $\mathrm{CP2}$   & \begin{tabular}{@{}l@{}}  $\mu_1^2=\mu_2^2$, $\lambda_1=\lambda_2$, $\lambda_3$, $\lambda_4$, $\lambda_5$, $\lambda_6=-\lambda_7$  \\    $\kappa_1=\kappa_2$, $\kappa_3=\kappa_4$, $\kappa_5=\kappa_6$, $\kappa_8=\kappa_9$, $\kappa_{11}=-\kappa_{12}$ \\ $\zeta_1=\zeta_2$, $\zeta_3$, $\zeta_4=\zeta_5$, $\zeta_6$, $\zeta_7=\zeta_8$, $\zeta_9$ \\$\zeta_{10}$, $\zeta_{11}=-\zeta_{12}$, $\zeta_{13}$, $\zeta_{14}=\zeta_{15}$, $\zeta_{16}$, $\zeta_{17}=-\zeta_{18}$, $\zeta_{19}=-\zeta_{20}$, $\zeta_{21}=-\zeta_{22}$ \end{tabular} &$D\geq4$ \\ \hline

$\mathrm{CP2}\otimes Z_2$   & \begin{tabular}{@{}l@{}}  $\mu_1^2=\mu_2^2$, $\lambda_1=\lambda_2$, $\lambda_3$, $\lambda_4$, $\lambda_5$  \\    $\kappa_1=\kappa_2$, $\kappa_3=\kappa_4$, $\kappa_5=\kappa_6$,  $\kappa_8=\kappa_9$  \\ $\zeta_1=\zeta_2$, $\zeta_3$, $\zeta_4=\zeta_5$, $\zeta_6$, $\zeta_7=\zeta_8$, $\zeta_9$, $\zeta_{10}$, $\zeta_{13}$, $\zeta_{14}=\zeta_{15}$, $\zeta_{16}$ \end{tabular} &$D\geq4$\\  \hline

    $\mathrm{CP3}$   & \begin{tabular}{@{}l@{}} $\mu_1=\mu_2$, $\lambda_1=\lambda_2$, $\lambda_3$, $\lambda_4$\\ $\kappa_1=\kappa_2$, $\kappa_3=\kappa_4$, $\kappa_5=\kappa_6$, $\kappa_7$\\ $\zeta_1=\zeta_2$, $\zeta_3$, $\zeta_4=\zeta_5$, $\zeta_6$, $\zeta_7=\zeta_8$, $\zeta_9$, $\zeta_{11}=\zeta_{12}$ \end{tabular} &$D\geq6$ \\ \hline

    $\mathrm{CP4}$   & \begin{tabular}{@{}l@{}}  $\mu_1=\mu_2$, $\lambda_1=\lambda_2$, $\lambda_3$, $\lambda_4$\\ $\kappa_1=\kappa_2$, $\kappa_3=\kappa_4$, $\kappa_5=\kappa_6$, $\kappa_8=-\kappa_9$\\ $\zeta_1=\zeta_2$, $\zeta_3$, $\zeta_4=\zeta_5$, $\zeta_6$, $\zeta_7=\zeta_8$, $\zeta_9$, $\zeta_{10}$, $\zeta_{14}=-\zeta_{15}$ \end{tabular} &$D\geq6$\\ \hline

 $\mathrm{CP2}\otimes Z_3$   & \begin{tabular}{@{}l@{}}  $\mu_1^2=\mu_2^2$, $\lambda_1=\lambda_2$, $\lambda_3$, $\lambda_4$  \\    $\kappa_1=\kappa_2$, $\kappa_3=\kappa_4$, $\kappa_5=\kappa_6$  \\ $\zeta_1=\zeta_2$, $\zeta_3$, $\zeta_4=\zeta_5$, $\zeta_6$, $\zeta_7=\zeta_8$, $\zeta_9$, $\zeta_{11}=-\zeta_{12}$ \end{tabular}&$D\geq8$ \\  \hline

  $\mathrm{CP2}\otimes Z_4$   & \begin{tabular}{@{}l@{}}  $\mu_1^2=\mu_2^2$, $\lambda_1=\lambda_2$, $\lambda_3$, $\lambda_4$  \\    $\kappa_1=\kappa_2$, $\kappa_3=\kappa_4$, $\kappa_5=\kappa_6$  \\ $\zeta_1=\zeta_2$, $\zeta_3$, $\zeta_4=\zeta_5$, $\zeta_6$, $\zeta_7=\zeta_8$, $\zeta_9$, $\zeta_{10}$  \end{tabular}&$D\geq8$ \\  \hline

\end{longtable*}
\caption{\it Parameter relations for the symmetric 2HDMEFT potentials, including dimension-6 and dimension-8 operators, under the possible discrete CP symmetries. The third column indicates what higher dimension operators must be included for the symmetry to be distinct.}
\label{tab:CP}
\end{table}
\addtocounter{table}{-1}

\section{Full Classification of Accidental Symmetries in 2HDMEFT}\label{sec:full}
 
The discrete symmetries can be combined with continuous symmetries to form additional symmetries. Note that some possible combinations result in unique parameter relations but they turn out to be equivalent through a change of basis. For example, one could construct a potential invariant under $S_2\otimes\mathrm{U(1)}_{\rm PQ}$ that has unique parameter relations. However, this symmetry group is isomorphic to $\mathrm{CP1}\otimes\mathrm{SO(2)}_{\rm HF}$. Hence, the resulting $S_2\otimes\mathrm{U(1)}_{\rm PQ}$- and $\mathrm{CP1}\otimes\mathrm{SO(2)}_{\rm HF}$-symmetric potentials are equivalent via a change of basis, and so they are not counted as distinct. 
We also require invariance under a reparameterisation of the Higgs doublets, via $\phi_i\to U_{ij}\phi_j$ with $U\in\mathrm{U(2)}$, which reduces many symmetric potentials to a reduced basis \cite{invariants}.  We omit symmetries that are equivalent under a $\mathrm{U(2)}$ basis transformation and keep the definitions of potentials in the most generic form. Hence, we omit the $\text{CP2}\otimes Z_n$ and $D_n$ symmetries and quote $\text{CP}n$ in the more general basis. Additionally, in the case where a discrete symmetry is equivalent to a continuous one, we quote the maximal continuous symmetry.
 
 The full list of accidental symmetries identified in the 2HDMEFT, with the presence of dimension-6 and dimension-8 operators, are given in Table \ref{tab:d8both}, along with the corresponding parameter relations. There are four additional symmetries that arise from the higher dimension discrete groups. Notably, none of these additional discrete groups forms a unique symmetry in combination with the continuous symmetries present in 2HDM. One can see that if non-renormalisable operators are neglected, we regain the 13 accidental symmetries reported in 2HDM \cite{ndarvishi,apilaftsis}, albeit with CP2 quoted in a more general basis. 

All parameter relations obtained in this section are also recovered utilising an alternative bilinear approach detailed in Appendix~\ref{ap:coadjoint}.

\begin{table}[t]
\begin{longtable}[h]{| s | s | t | s |}
\hline
\small{No.}  &  \small{Symmetry}  &  \small{Non-zero parameters of Symmetric 2HDMEFT Potential} & \small{Dim} \\
\hline \hline

1& $\mathrm{CP1}$ & \begin{tabular}{@{}l@{}}  $\mu_1^2$, $\mu_2^2$, Re($m_{12}^2$), $\lambda_1$, $\lambda_2$, $\lambda_3$, $\lambda_4$, Re($\lambda_5$), Re($\lambda_6$), Re($\lambda_7$)  \\  $\kappa_1,\dots,\,\kappa_6$, Re($\kappa_7,\dots,\,\kappa_{13}$) \\ $\zeta_1,\dots,\,\zeta_9$, Re($\zeta_{10},\dots,\,\zeta_{22}$)
 \end{tabular} &$D\geq4$  \\ \hline

2& $Z_2$ &  \begin{tabular}{@{}l@{}} $\mu_1^2$, $\mu_2^2$, $\lambda_1$, $\lambda_2$, $\lambda_3$, $\lambda_4$,  $\lambda_5$    \\ $\kappa_1$, $\kappa_2$, $\kappa_3$, $\kappa_4$, $\kappa_5$, $\kappa_6$, $\kappa_8$, $\kappa_9$ \\$\zeta_1$, $\zeta_2$, $\zeta_3$, $\zeta_4$, $\zeta_5$ $\zeta_6$, $\zeta_7$, $\zeta_8$, $\zeta_9$, $\zeta_{10}$, $\zeta_{13}$, $\zeta_{14}$, $\zeta_{15}$, $\zeta_{16}$ \end{tabular} & $D\geq4$  \\ \hline

3& $Z_3$ & \begin{tabular}{@{}l@{}}  $\mu_1^2$, $\mu_2^2$, $\lambda_1$, $\lambda_2$, $\lambda_3$, $\lambda_4$  \\ $\kappa_1$, $\kappa_2$, $\kappa_3$, $\kappa_4$, $\kappa_5$, $\kappa_6$, $\kappa_7$ \\$\zeta_1$, $\zeta_2$, $\zeta_3$, $\zeta_4$, $\zeta_5$, $\zeta_6$, $\zeta_7$, $\zeta_8$, $\zeta_9$, $\zeta_{11}$, $\zeta_{12}$ \end{tabular} & $D\geq6$  \\ \hline

4& $Z_4$ & \begin{tabular}{@{}l@{}}  $\mu_1^2$, $\mu_2^2$, $\lambda_1$, $\lambda_2$, $\lambda_3$, $\lambda_4$  \\ $\kappa_1$, $\kappa_2$, $\kappa_3$, $\kappa_4$, $\kappa_5$, $\kappa_6$ \\ $\zeta_1$, $\zeta_2$, $\zeta_3$, $\zeta_4$, $\zeta_5$, $\zeta_6$, $\zeta_7$, $\zeta_8$, $\zeta_9$, $\zeta_{10}$ \end{tabular} & $D\geq8$  \\ \hline

5&  $\mathrm{CP2}$   & \begin{tabular}{@{}l@{}}  $\mu_1^2=\mu_2^2$, $\lambda_1=\lambda_2$, $\lambda_3$, $\lambda_4$, $\lambda_5$, $\lambda_6=-\lambda_7$  \\    $\kappa_1=\kappa_2$, $\kappa_3=\kappa_4$, $\kappa_5=\kappa_6$, $\kappa_8=\kappa_9$, $\kappa_{11}=-\kappa_{12}$ \\ $\zeta_1=\zeta_2$, $\zeta_3$, $\zeta_4=\zeta_5$, $\zeta_6$, $\zeta_7=\zeta_8$, $\zeta_9$, \\ $\zeta_{10}$, $\zeta_{11}=-\zeta_{12}$, $\zeta_{13}$, $\zeta_{14}=\zeta_{15}$, $\zeta_{16}$, $\zeta_{17}=-\zeta_{18}$, $\zeta_{19}=-\zeta_{20}$, $\zeta_{21}=-\zeta_{22}$ \end{tabular} & $D\geq4$ \\ \hline

  6&   $\mathrm{CP3}$   & \begin{tabular}{@{}l@{}} $\mu_1=\mu_2$, $\lambda_1=\lambda_2$, $\lambda_3$, $\lambda_4$\\ $\kappa_1=\kappa_2$, $\kappa_3=\kappa_4$, $\kappa_5=\kappa_6$, $\kappa_7$\\ $\zeta_1=\zeta_2$, $\zeta_3$, $\zeta_4=\zeta_5$, $\zeta_6$, $\zeta_7=\zeta_8$, $\zeta_9$, $\zeta_{11}=\zeta_{12}$  \end{tabular} & $D\geq6$  \\ \hline

 7&   $\mathrm{CP4}$   & \begin{tabular}{@{}l@{}}  $\mu_1=\mu_2$, $\lambda_1=\lambda_2$, $\lambda_3$, $\lambda_4$ \\ $\kappa_1=\kappa_2$, $\kappa_3=\kappa_4$, $\kappa_5=\kappa_6$, $\kappa_8=-\kappa_9$ \\ $\zeta_1=\zeta_2$, $\zeta_3$, $\zeta_4=\zeta_5$, $\zeta_6$, $\zeta_7=\zeta_8$, $\zeta_9$, $\zeta_{10}$, $\zeta_{14}=-\zeta_{15}$ \end{tabular} & $D\geq6$  \\ \hline




8 & $\mathrm{U(1)_{PQ}}$ & \begin{tabular}{@{}l@{}} $\mu_1^2$, $\mu_2^2$, $\lambda_1$, $\lambda_2$, $\lambda_3$, $\lambda_4$\\ $\kappa_1$, $\kappa_2$, $\kappa_3$, $\kappa_4$, $\kappa_5$, $\kappa_6$\\ $\zeta_1$, $\zeta_2$, $\zeta_3$, $\zeta_4$, $\zeta_5$, $\zeta_6$, $\zeta_7$, $\zeta_8$, $\zeta_9$ \end{tabular} & $D\geq4$  \\ \hline


9 & $\mathrm{CP1}\otimes\mathrm{SO(2)_{\rm HF}}$ &  \begin{tabular}{@{}l@{}}$\mu_1^2=\mu_2^2$, $\lambda_1=\lambda_2$, $\lambda_3$, $\lambda_4$, Re($\lambda_5)=2\lambda_1-\lambda_{34}$\\ $\kappa_1=\kappa_2$, $\kappa_3=\kappa_4$, $\kappa_5=\kappa_6$, Re$(\kappa_8)=$Re$(\kappa_9)=\frac{1}{2}(3\kappa_1-\kappa_3-\kappa_5)$ \\$\zeta_1=\zeta_2$, $\zeta_3$, $\zeta_4=\zeta_5$, $\zeta_6$, $\zeta_7=\zeta_8$, $\zeta_9$, \\
$\mathrm{Re}(\zeta_{10})=-\frac{1}{4}\Re(\zeta_{13})+\frac{1}{2}\Re(\zeta_{14})-\frac{1}{4}\Re(\zeta_{16})$, $\mathrm{Re}(\zeta_{13})=\frac{1}{6}(4\zeta_1+2\zeta_3-4\zeta_4-4\zeta_6+2\zeta_7-\zeta_9)$, \\ $\mathrm{Re}(\zeta_{14})=\mathrm{Re}(\zeta_{15})=\frac{1}{2}(4\zeta_1-\zeta_4-\zeta_7)$,
$\mathrm{Re}(\zeta_{16})=\frac{1}{2}(4\zeta_1-2\zeta_3+2\zeta_4-\zeta_9)$
\end{tabular}  & $D\geq4$   \\ \hline

10 & $\mathrm{SU(2)_{HF}}$ & \begin{tabular}{@{}l@{}} $\mu_1^2=\mu_2^2$, $\lambda_1=\lambda_2$, $\lambda_3$, $\lambda_4=2\lambda_1-\lambda_3$ \\$\kappa_1=\kappa_2$, $\kappa_3=\kappa_4$, $\kappa_5=\kappa_6=3\kappa_1-\kappa_3$ \\ $\zeta_1=\zeta_2$, $\zeta_3$, $\zeta_4=\zeta_5$, $\zeta_6=2\zeta_1+\zeta_3-2\zeta_4$, $\zeta_7=\zeta_8=4\zeta_1-\zeta_4$, $\zeta_9=4\zeta_1-2\zeta_3+2\zeta_4$  \end{tabular} & $D\geq4$  \\ \hline \hline

11 & $\mathrm{Sp(2)}_{\phi_1+\phi_2}$ & \begin{tabular}{@{}l@{}}  $\mu_1^2$, $\mu_2^2$, Re($m_{12}^2$), $\lambda_1$, $\lambda_2$, $\lambda_3$, $\lambda_4=\text{Re}(\lambda_5)$, Re($\lambda_6$), Re($\lambda_7$) \\ $\kappa_1$, $\kappa_2$, $\kappa_3$, $\kappa_4$, $\kappa_5=2\mathrm{Re}(\kappa_8)$, $\kappa_6=2\mathrm{Re}(\kappa_9)$, $\mathrm{Re}(\kappa_7)=\frac{1}{3}\mathrm{Re}(\kappa_{10})$, Re($\kappa_{11}$), Re($\kappa_{12}$), Re($\kappa_{13}$) \\ 
$\zeta_1$, $\zeta_2$, $\zeta_3$, $\zeta_4$, $\zeta_5$, $\zeta_6=6\mathrm{Re}(\zeta_{10})=\frac{3}{2}\mathrm{Re}(\zeta_{13})$,
$\zeta_7=2\Re(\zeta_{14})$, $\zeta_8=2\Re(\zeta_{15})$, $\zeta_9=2\Re(\zeta_{16})$,  \\ $\Re(\zeta_{11})=\frac{1}{3}\Re(\zeta_{17})$, $\Re(\zeta_{12})=\frac{1}{3}\Re(\zeta_{18})$, $\Re(\zeta_{19})$, $\Re(\zeta_{20})$, $\Re(\zeta_{21})$, $\Re(\zeta_{22})$ 
\end{tabular}  & $D\geq4$   \\ \hline

 

12 & $S_2\otimes\mathrm{Sp(2)}_{\phi_1+\phi_2}$&  \begin{tabular}{@{}l@{}} $\mu_1^2=\mu_2^2$,  Re($m_{12}^2$), $\lambda_1=\lambda_2$, $\lambda_3$, $\lambda_4=\text{Re}(\lambda_5)$, Re($\lambda_6$)=Re($\lambda_7$) \\ $\kappa_1=\kappa_2$, $\kappa_3=\kappa_4$, $\kappa_5=\kappa_6=2\mathrm{Re}(\kappa_8)=2\mathrm{Re}(\kappa_9)$, $\mathrm{Re}(\kappa_7)=\frac{1}{3}\mathrm{Re}(\kappa_{10})$, Re($\kappa_{11})=\mathrm{Re}(\kappa_{12}$), Re($\kappa_{13}$) \\ $\zeta_1=\zeta_2$, $\zeta_3$, $\zeta_4=\zeta_5$, $\zeta_6=6\mathrm{Re}(\zeta_{10})=\frac{3}{2}\mathrm{Re}(\zeta_{13})$, $\zeta_7=\zeta_8=2\Re(\zeta_{14})=2\Re(\zeta_{15})$, $\zeta_9=2\Re(\zeta_{16})$ \\ $\Re(\zeta_{11})=\Re(\zeta_{12})=\frac{1}{3}\Re(\zeta_{17})=\frac{1}{3}\Re(\zeta_{18})$, $\Re(\zeta_{19})=\Re(\zeta_{20})$, $\Re(\zeta_{21})=\Re(\zeta_{22})$ \end{tabular}  & $D\geq4$     \\ \hline

13 & $\mathrm{CP2}\otimes\mathrm{Sp(2)}_{\phi_1+\phi_2}$ &    \begin{tabular}{@{}l@{}} $\mu_1^2=\mu_2^2$, $\lambda_1=\lambda_2$, $\lambda_3$, $\lambda_4=\mathrm{Re}(\lambda_5)$, $\Re(\lambda_6)=-\Re(\lambda_7)$ \\ $\kappa_1=\kappa_2$, $\kappa_3=\kappa_4$, $\kappa_5=\kappa_6=2\mathrm{Re}(\kappa_8)=2\mathrm{Re}(\kappa_9)$, $\Re(\kappa_{11})=-\Re(\kappa_{12})$ \\ $\zeta_1=\zeta_2$, $\zeta_3$, $\zeta_4=\zeta_5$, $\zeta_6=6\mathrm{Re}(\zeta_{10})=\frac{3}{2}\mathrm{Re}(\zeta_{13})$, $\zeta_7=\zeta_8=2\Re(\zeta_{14})=2\Re(\zeta_{15})$, $\zeta_9=2\Re(\zeta_{16})$\\ $\Re(\zeta_{11})=-\Re(\zeta_{12})=\frac{1}{3}\Re(\zeta_{17})=-\frac{1}{3}\Re(\zeta_{18})$,  $\Re(\zeta_{19})=-\Re(\zeta_{20})$, $\Re(\zeta_{21})=-\Re(\zeta_{22})$   \end{tabular}  & $D\geq4$   \\ \hline



14 & $\mathrm{U(1)_{PQ}}\otimes\mathrm{Sp(2)}_{\phi_1\phi_2}$ & \begin{tabular}{@{}l@{}}  $\mu_1^2=\mu_2^2$, $\lambda_1=\lambda_2=\frac{1}{2}\lambda_3$, $\lambda_4$\\ $\kappa_1=\kappa_2=\frac{1}{3}\kappa_3=\frac{1}{3}\kappa_4$, $\kappa_5=\kappa_6$ \\$\zeta_1=\zeta_2=\frac{1}{6}\zeta_3=\frac{1}{4}\zeta_4=\frac{1}{4}\zeta_5$, $\zeta_6$, $\zeta_7=\zeta_8=\frac{1}{2}\zeta_9$ \end{tabular} & $D\geq4$  \\ \hline

15 & $\mathrm{Sp(2)}_{\phi_1}\otimes\mathrm{Sp(2)}_{\phi_2}$ & \begin{tabular}{@{}l@{}}  $\mu_1^2$, $\mu_2^2$, $\lambda_1$, $\lambda_2$, $\lambda_3$\\ $\kappa_1$, $\kappa_2$, $\kappa_3$, $\kappa_4$\\ $\zeta_1$, $\zeta_2$, $\zeta_3$, $\zeta_4$, $\zeta_5$     \end{tabular} & $D\geq4$   \\ \hline

16 & $S_2\otimes\mathrm{Sp(2)}_{\phi_1}\otimes\mathrm{Sp(2)}_{\phi_2}$ &  \begin{tabular}{@{}l@{}} $\mu_1^2=\mu_2^2$, $\lambda_1=\lambda_2$, $\lambda_3$\\  $\kappa_1=\kappa_2$, $\kappa_3=\kappa_4$\\   $\zeta_1=\zeta_2$, $\zeta_3$, $\zeta_4=\zeta_5$  \end{tabular} & $D\geq4$  \\ \hline

17 & $\mathrm{Sp(4)}$ & \begin{tabular}{@{}l@{}}  $\mu_1^2=\mu_2^2$, $\lambda_1=\lambda_2={1\over 2}\lambda_3$ \\$\kappa_1=\kappa_2=\frac{1}{3}\kappa_3=\frac{1}{3}\kappa_4$ \\$\zeta_1=\zeta_2=\frac{1}{6}\zeta_3=\frac{1}{4}\zeta_4=\frac{1}{4}\zeta_5$ \end{tabular} & $D\geq4$   \\ \hline

\end{longtable}

\caption{\it Parameter relations for the symmetric 2HDMEFT potentials, including dimension-6 and dimension-8 operators, under the possible accidental symmetries. The 4th column indicates what higher dimension operators must be included for the symmetry to be distinct.  Note that the last 7 entries which contain the symplectic group are only custodially symmetric.}
\label{tab:d8both}
\end{table}

\addtocounter{table}{-1}

\section{Conclusions}\label{sec:con}

We have shown that additional accidental symmetries arise when higher-order operators are included in the 2HDMEFT potential. These additional symmetries only appear from higher dimensional discrete symmetry groups. Taking into consideration all possible continuous, discrete and combined symmetries that may occur in the 2HDMEFT potential, we have been able to identify the full list of 17 accidental symmetries in the 2HDMEFT that includes dimension-6 and dimension-8 operators. By employing an approach based on prime invariants, we have also derived the relations that govern the theoretical parameters of the corresponding accidentally symmetric effective potentials. When non-renormalisable terms are omitted from the potential, we recover the usual 13 accidental symmetries of the~2HDM. 

Following an alternative approach as detailed in Appendix~\ref{ap:coadjoint}, the above 17 symmetries have also been obtained by making use of symmetry group generators in the bi-adjoint representation. With the help of the same approach, we have also provided a systematic algebraic 
method that enables one to obtain the parameter relations of accidentally symmetric 2HDMEFT potentials of dimension~$2n$, once the symmetry generators are specified.

The phenomenological implications of the operators and the accidental symmetries discussed${}$ here lie beyond the scope of the present work. Nonetheless, we anticipate that the higher-order operators arising from 2HDMEFT induce corrections to the Higgs self-couplings and Yukawa interactions in the 2HDM. The Higgs self-interactions may be probed by double Higgs production at the LHC or future high-energy colliders, which may have the capability to provide constraints on dimension-6 operators. On the other hand, the constraints on Yukawa interactions look more promising and may be probed by measurements that involve, for instance, the production of four quarks. We plan to return to such dedicated phenomenological investigations in the near future. 

\vspace{-5mm}
\subsection*{Acknowledgements}
\vspace{-3mm}
\noindent
The work of AP and ND is supported in part by the Lancaster–-Manchester–-Sheffield Consortium\- for Fundamental Physics, under STFC research grant ST/P000800/1. The work of YP is supported partially by the ERC research grant 817719-TheHiggsAndThe7Tops.

\vfill\eject

\appendix
\section{2HDM and 2HDMEFT Potentials}\label{ap:V}

In Section \ref{sec:s2}, we have given the 2HDMEFT potential, $V_\text{2HDMEFT}$, by explicitly writing down the dimension-6 and dimension-8 operators and neglecting higher orders,
 \begin{equation}
V_\text{2HDMEFT}=V_\mathrm{2HDM}+V^\mathrm{(dim=6)}_\mathrm{2HDM}+V^\mathrm{(dim=8)}_\mathrm{2HDM}+\mathcal{O}(\Lambda^{-6})\,.
 \end{equation}
In the Higgs doublet field space, the most general renormalisable 2HDM potential may be conventionally expressed as
\begin{align}\nonumber
V_\mathrm{2HDM}=&-\mu^2_1(\phi^{\dagger}_1\phi_1)-\mu^2_2(\phi^{\dagger}_2\phi_2)-\Big[m_{12}^2(\phi_1^{\dagger}\phi_2)+\mathrm{H.c}\Big]\\\nonumber
&+\lambda_1(\phi^{\dagger}_1\phi_1)^2+\lambda_2(\phi^{\dagger}_2\phi_2)^2+\lambda_3(\phi^{\dagger}_1\phi_1)(\phi^{\dagger}_2\phi_2)+\lambda_4(\phi^{\dagger}_1\phi_2)(\phi^{\dagger}_2\phi_1)\\
&+\bigg[\,\frac{1}{2}\lambda_5(\phi^{\dagger}_1\phi_2)^2+\lambda_6(\phi^{\dagger}_1\phi_1)(\phi^{\dagger}_1\phi_2)+\lambda_7(\phi^{\dagger}_2\phi_2)(\phi^{\dagger}_1\phi_2)+\mathrm{H.c.}\bigg]\,.
\end{align}
In addition, the  dimension-6 and dimension-8 2HDMEFT potentials are
 \begin{align}
V^\mathrm{(dim=6)}_\mathrm{2HDM}=
 \frac{1}{\Lambda^2}\bigg\{ &\kappa_1 ( \phi_1^{\dagger} \phi_1)^3
 + \kappa_2 ( \phi_2^{\dagger} \phi_2)^3
 + \kappa_3 ( \phi_1^{\dagger} \phi_1)^2( \phi_2^{\dagger} \phi_2)
  + \kappa_4 ( \phi_1^{\dagger} \phi_1)( \phi_2^{\dagger} \phi_2)^2
   \nonumber \\&
 + \kappa_5 ( \phi_1^{\dagger} \phi_2)( \phi_2^{\dagger} \phi_1)  ( \phi_1^{\dagger} \phi_1)
+ \kappa_6 ( \phi_1^{\dagger} \phi_2)( \phi_2^{\dagger} \phi_1)  ( \phi_2^{\dagger} \phi_2)
  \nonumber \\& 
 + \left[\, 
 \kappa_7 ( \phi_1^{\dagger} \phi_2)^3
 + \kappa_8 ( \phi_1^{\dagger} \phi_2)^2  ( \phi_1^{\dagger} \phi_1)
 + \kappa_9 ( \phi_1^{\dagger} \phi_2)^2 ( \phi_2^{\dagger} \phi_2) \right.
 \nonumber \\&
\left.  +  \kappa_{10} ( \phi_1^{\dagger} \phi_2)^2 ( \phi_2^{\dagger} \phi_1)
  +  \kappa_{11} ( \phi_1^{\dagger} \phi_2) ( \phi_1^{\dagger} \phi_1)^2
   +  \kappa_{12} ( \phi_1^{\dagger} \phi_2) ( \phi_2^{\dagger} \phi_2)^2
    \right.
 \nonumber \\&
\left.  
    +  \kappa_{13}  ( \phi_1^{\dagger} \phi_2)  ( \phi_1^{\dagger} \phi_1) ( \phi_2^{\dagger} \phi_2)
+{\rm H.c.} \right]\bigg\}\,,
\\
    \nonumber
V^\mathrm{(dim=8)}_\mathrm{2HDM}=\frac{1}{\Lambda^4}\bigg\{&\zeta_1(\phi_1^\dagger\phi_1)^4+\zeta_2(\phi_2^\dagger\phi_2)^4+\zeta_3(\phi_1^\dagger\phi_1)^2(\phi_2^\dagger\phi_2)^2+\zeta_4(\phi_1^\dagger\phi_1)(\phi_2^\dagger\phi_2)^3\\\nonumber&+\zeta_5(\phi_1^\dagger\phi_1)^3(\phi_2^\dagger\phi_2)+\zeta_6(\phi_1^\dagger\phi_2)^2(\phi_2^\dagger\phi_1)^2+\zeta_7(\phi_1^\dagger\phi_2)(\phi_2^\dagger\phi_1)(\phi_1^\dagger\phi_1)^2\\\nonumber&+\zeta_8(\phi_1^\dagger\phi_2)(\phi_2^\dagger\phi_1)(\phi_2^\dagger\phi_2)^2+\zeta_9(\phi_1^\dagger\phi_2)(\phi_2^\dagger\phi_1)(\phi_1^\dagger\phi_1)(\phi_2^\dagger\phi_2) \\\nonumber&+\Big[
\zeta_{10}(\phi_1^\dagger\phi_2)^4+\zeta_{11}(\phi_1^\dagger\phi_2)^3(\phi_1^\dagger\phi_1)+\zeta_{12}(\phi_1^\dagger\phi_2)^3(\phi_2^\dagger\phi_2)\\\nonumber&+\zeta_{13}(\phi_1^\dagger\phi_2)^3(\phi_2^\dagger\phi_1)+\zeta_{14}(\phi_1^\dagger\phi_2)^2(\phi_1^\dagger\phi_1)^2+\zeta_{15}(\phi_1^\dagger\phi_2)^2(\phi_2^\dagger\phi_2)^2\\\nonumber&+\zeta_{16}(\phi_1^\dagger\phi_2)^2(\phi_1^\dagger\phi_1)(\phi_2^\dagger\phi_2)+\zeta_{17}(\phi_1^\dagger\phi_2)^2(\phi_1^\dagger\phi_1)(\phi_2^\dagger\phi_1)\\\nonumber&+\zeta_{18}(\phi_1^\dagger\phi_2)^2(\phi_2^\dagger\phi_2)(\phi_2^\dagger\phi_1)+\zeta_{19}(\phi_1^\dagger\phi_2)(\phi_1^\dagger\phi_1)^3+\zeta_{20}(\phi_1^\dagger\phi_2)(\phi_2^\dagger\phi_2)^3\\&+\zeta_{21}(\phi_1^\dagger\phi_2)(\phi_1^\dagger\phi_1)^2(\phi_2^\dagger\phi_2)+\zeta_{22}(\phi_1^\dagger\phi_2)(\phi_1^\dagger\phi_1)(\phi_2^\dagger\phi_2)^2+\mathrm{H.c.}
    \Big]\bigg\}\,.
\end{align} 

\section{Free Parameters in Dimension-$2n$ 2HDMEFT Operators}\label{ap:nparams}
 
In the 2HDMEFT framework, operators of higher order dimension~$2n$ (with $n>2$) can be 
concisely written down in the bilinear representation via a symmetric rank-$n$ tensor, $\Gamma^{(\mathrm{dim}=2n)}$, as
\begin{equation}
     V^{(\mathrm{dim}=2n)}_\mathrm{2HDM}=\ \frac{1}{\Lambda^{(2n-4)}}\Gamma^{(\mathrm{dim}=2n)}_{A_1,A_2\ldots, A_n}R^{A_1}R^{A_2}\ldots R^{A_n}\,.
 \end{equation}
We may now infer the number of free parameters in $V^{(\mathrm{dim}=2n)}_\mathrm{2HDM}$ by counting the number of independent entries of the symmetric rank-$n$ tensor $\Gamma^{(\mathrm{dim}=2n)}$. For a general symmetric rank-$n$ tensor of dimension $k$ (each index running from $1$ to $k$), the number of independent entries can be determined by the binomial coefficient,
\begin{equation}
    \begin{pmatrix} n+k-1 \\ n \end{pmatrix}=\frac{(n+k-1)!}{n!(k-1)!}\,.
\end{equation}
In the $R^A$-space, terms involving $R^{4,5}$ can be set to zero to preserve $\mathrm{U(1)}_Y$ symmetry.  As such, the number of free parameters in $V^{(\mathrm{dim}=2n)}_\mathrm{2HDM}$ is given by the number of independent components of $\Gamma^{(\mathrm{dim}=2n)}_{\nu_1,\nu_2,\ldots,\nu_n}$ with $\nu_i=0,1,2,3$. Hence, for a $\mathrm{U(1)}_Y$-symmetric model, the number of couplings $V^{(\mathrm{dim}=2n)}_\mathrm{2HDM}$ is given by
\begin{equation}
    N^{(\mathrm{dim}=2n)}=\begin{pmatrix} 2n+3 \\ 2n \end{pmatrix}=\frac{(2n+3)!}{3! (2n)!}=\frac{(2n+1)(2n+2)(2n+3)}{6}.
\end{equation}

We can then compute the additional free parameters in the $\mathrm{U(1)}_Y$-symmetric 2HDMEFT up to any dimension $2n$. In Table~\ref{tab:nparams}, we show the number of
independent theoretical parameters up to dimension 20.

\begin{table}[h]
     \begin{longtable}{|s|s|s|}
     \hline
     Dimension   &  No. Parameters, $N^{(\mathrm{dim}=2n)}$  &   Cumulative No. Parameters, $N^{(\mathrm{dim}\leq2n)}$ \\\hline\hline
     $2n\leq4$   &   14           &   14  \\
     $2n=6$      &   20              &   34  \\
     $2n=8$      &   35              &   69  \\
     $2n=10$     &   56              &   125 \\
     $2n=12$     &   84              &   209 \\
          $2n=14$     &   120             &   329 \\
     $2n=16$     &   165             &   494 \\
     $2n=18$     &   220             &   714 \\
     $2n=20$     &   286             &   1000\\
     \hline
     \end{longtable}
     \caption{ \textit{Number of free parameters arising from 2HDMEFT higher order operators $V^{(\mathrm{dim}=2n)}_\mathrm{2HDM}$ and number of free parameter in 2HDMEFT potentials up to dimension $2n$. }}
     \label{tab:nparams}
\end{table}

\section{2HDMEFT Couplings in the Bilinear Formalism}\label{ap:KZ}

In Section \ref{sec:s2}, we have seen that in this bilinear representation, the additional higher order terms of the 2HDMEFT potential may be written as
 \begin{align}
     V_\mathrm{2HDM}^\mathrm{(dim=6)} =&\ \frac{1}{\Lambda^2}K_{ABC}\,R^AR^BR^C\,,\\
     V^\mathrm{(dim=8)}_\mathrm{2HDM}=&\ \frac{1}{\Lambda^4}Z_{ABCD}\,R^AR^BR^CR^D,
 \end{align}
where $K_{ABC}$ and $Z_{ABCD}$ are symmetric rank-3 and rank-4 tensors describing the hexic and octic couplings, respectively. 

For the dimension-6 part of the 2HDMEFT potential, we find that the elements of the symmetric tensor $K_{ABC}$ are
{\tiny \begin{flalign*}
    &K_{0,0,0}= \frac{1}{8}\left(\kappa_1+\kappa_2+\kappa_3+\kappa_4\right), \quad
    &&K_{0,0,1}= \frac{1}{12}\left(\Re\left[\kappa_{11}\right]+\Re\left[\kappa_{12}\right]+\Re\left[\kappa_{13}\right]\right),&
    \nonumber \\
    &K_{0,0,2}= \frac{1}{12} \left(-\Im\left[\kappa_{11}\right]-\Im\left[\kappa_{12}\right]-\Im\left[\kappa_{13}\right]\right),
    &&K_{0,0,3}= \frac{1}{24} \left(3 \kappa_1-3 \kappa_2+\kappa_3-\kappa_4\right),
    \nonumber \\
    &K_{0,1,1}= \frac{1}{24} \left(\kappa_5+\kappa_6+2 \Re\left[\kappa_8\right]+2 \Re\left[\kappa_9\right]\right),
    &&K_{0,1,2}= \frac{1}{12} \left(-\Im\left[\kappa_8\right]-\Im\left[\kappa_9\right]\right),
    \nonumber \\
    &K_{0,1,3}= \frac{1}{12} \left(\Re\left[\kappa_{11}\right]-\Re\left[\kappa_{12}\right]\right),
    &&K_{0,2,2}= \frac{1}{24} \left(\kappa_5+\kappa_6-2 \Re\left[\kappa_8\right]-2 \Re\left[\kappa_9\right]\right),
    \nonumber \\
    &K_{0,2,3}= \frac{1}{12} \left(\Im\left[\kappa_{12}\right]-\Im\left[\kappa_{11}\right]\right),
    &&K_{0,3,3}= \frac{1}{24} \left(3 \kappa_1+3 \kappa_2-\kappa_3-\kappa_4\right),
    \nonumber \\
    &K_{1,1,1}= \frac{1}{4} \left(\Re\left[\kappa_7\right]+\Re\left[\kappa_{10}\right]\right),
    &&K_{1,1,2}= \frac{1}{12} \left(-3\Im\left[\kappa_7\right]-\Im\left[\kappa_{10}\right]\right),
    \nonumber \\
    &K_{1,1,3}= \frac{1}{24} \left(\kappa_5-\kappa_6+2 \Re\left[\kappa_8\right]-2 \Re\left[\kappa_9\right]\right),
    &&K_{1,2,2}= \frac{1}{12} \left(\Re\left[\kappa_{10}\right]-3 \Re\left[\kappa_7\right]\right),
    \nonumber \\
    &K_{1,2,3}= \frac{1}{12} \left(\Im\left[\kappa_9\right]-\Im\left[\kappa_8\right]\right),
    &&K_{1,3,3}= \frac{1}{12} \left(\Re\left[\kappa_{11}\right]+\Re\left[\kappa_{12}\right]-\Re\left[\kappa_{13}\right]\right),
    \nonumber \\
    &K_{2,2,2}= \frac{1}{4} \left(\Im\left[\kappa_7\right]-\Im\left[\kappa_{10}\right]\right),
    &&K_{2,2,3}= \frac{1}{24} \left(\kappa_5-\kappa_6-2 \Re\left[\kappa_8\right]+2 \Re\left[\kappa_9\right]\right),
    \nonumber \\
    &K_{2,3,3}= \frac{1}{12} \left(-\Im\left[\kappa_{11}\right]-\Im\left[\kappa_{12}\right]+\Im\left[\kappa_{13}\right]\right),
    &&K_{3,3,3}= \frac{1}{8} \left(\kappa_1-\kappa_2-\kappa_3+\kappa_4\right).
    \nonumber
\end{flalign*}
}

Correspondingly, for the dimension-8 part of the 2HDMEFT potential, the elements of the symmetric tensor $Z_{ABCD}$ are given by
{\tiny \begin{flalign*}
  &Z_ {0,0,0,0}= \frac{1}{16} \left(\zeta_1+\zeta_2+\zeta_3+\zeta_4+\zeta_5\right), \qquad
  &&Z_ {0,0,0,1}= \frac{1}{32} \left[\Re\left[\zeta_{19}\right]+\Re\left[\zeta_{20}\right]+\Re\left[\zeta_{21}\right]+\Re\left[\zeta_{22}\right]\right],\\ \nonumber 
  &Z_ {0,0,0,2}= \frac{1}{32} \left(-\Im\left[\zeta_{19}\right]-\Im\left[\zeta_{20}\right]-\Im\left[\zeta_{21}\right]-\Im\left[\zeta_{22}\right]\right), \qquad
  &&Z_ {0,0,0,3}=  \frac{1}{32} \left(2 \zeta_1-2\zeta_2-\zeta_4+\zeta_5\right),\\ \nonumber 
  &Z_ {0,0,1,1}=  \frac{1}{96} \left(2 \Re\left[\zeta_{14}\right]+2\Re\left[\zeta_{15}\right]+2 \Re\left[\zeta_{16}\right]+\zeta_7+\zeta_8+\zeta_9\right), \qquad
  &&Z_ {0,0,1,2}=  \frac{1}{48} \left(-\Im\left[\zeta_{14}\right]-\Im\left[\zeta_{15}\right]-\Im\left[\zeta_{16}\right]\right),\\ \nonumber 
  &Z_ {0,0,1,3}=  \frac{1}{96} \left(3 \Re\left[\zeta_{19}\right]-3\Re\left[\zeta_{20}\right]+\Re\left[\zeta_{21}\right]-\Re\left[\zeta_{22}\right]\right), \qquad
  &&Z_ {0,0,2,2}=  \frac{1}{96} \left(-2 \Re\left[\zeta_{14}\right]-2\Re\left[\zeta_{15}\right]-2 \Re\left[\zeta_{16}\right]+\zeta_7+\zeta_8+\zeta_9\right),\\ \nonumber 
  &Z_ {0,0,2,3}=  \frac{1}{96} \left(-3 \Im\left[\zeta_{19}\right]+3\Im\left[\zeta_{20}\right]-\Im\left[\zeta_{21}\right]+\Im\left[\zeta_{22}\right]\right), \qquad
  &&Z_ {0,0,3,3}=  \frac{1}{48} \left(3 \zeta_1+3 \zeta_2-\zeta_3\right),\\ \nonumber 
  &Z_ {0,1,1,1}=  \frac{1}{32} \left(\Re\left[\zeta_{11}\right]+\Re\left[\zeta_{12}\right]+\Re\left[\zeta_{17}\right]+\Re\left[\zeta_{18}\right]\right), \qquad
  &&Z_ {0,1,1,2}=  \frac{1}{96} \left(-3\Im\left[\zeta_{11}\right]-3\Im\left[\zeta_{12}\right]-\Im\left[\zeta_{17}\right]-\Im\left[\zeta_{18}\right]\right),\\ \nonumber 
  &Z_ {0,1,1,3}=  \frac{1}{96} \left(2 \Re\left[\zeta_{14}\right]-2 \Re\left[\zeta_{15}\right]+\zeta_7-\zeta_8\right), \qquad
  &&Z_ {0,1,2,2}=  \frac{1}{96} \left(-3\Re\left[\zeta_{11}\right]-3\Re\left[\zeta_{12}\right]+\Re\left[\zeta_{17}\right]+\Re\left[\zeta_{18}\right]\right), \\\nonumber
  &Z_ {0,1,2,3}=  \frac{1}{48} \left(\Im\left[\zeta_{15}\right]-\Im\left[\zeta_{14}\right]\right), \qquad
  &&Z_ {0,1,3,3}=  \frac{1}{96} \left(3\Re\left[\zeta_{19}\right]+3\Re\left[\zeta_{20}\right]-\Re\left[\zeta_{21}\right]-\Re\left[\zeta_{22}\right]\right), \\\nonumber
  &Z_ {0,2,2,2}=  \frac{1}{32} \left(\Im\left[\zeta_{11}\right]+\Im\left[\zeta_{12}\right]-\Im\left[\zeta_{17}\right]-\Im\left[\zeta_{18}\right]\right), \qquad
  &&Z_ {0,2,2,3}=  \frac{1}{96} \left(-2 \Re\left[\zeta_{14}\right]+2\Re\left[\zeta_{15}\right]+\zeta_7-\zeta_8\right), \\ \nonumber 
  &Z_ {0,2,3,3}=  \frac{1}{96} \left(-3 \Im\left[\zeta_{19}\right]-3\Im\left[\zeta_{20}\right]+\Im\left[\zeta_{21}\right]+\Im\left[\zeta_{22}\right]\right), \qquad 
  &&Z_ {0,3,3,3}=  \frac{1}{32} \left(2 \zeta_1-2 \zeta_2+\zeta_4-\zeta_5\right), \\ \nonumber 
  &Z_ {1,1,1,1}=  \frac{1}{16} \left(2 \Re\left[\zeta_{10}\right]+2\Re\left[\zeta_{13}\right]+\zeta_6\right), \qquad
  &&Z_ {1,1,1,2}=  \frac{1}{16} \left(-2 \Im\left[\zeta_{10}\right]-\Im\left[\zeta_{13}\right]\right), \\ \nonumber 
  &Z_ {1,1,1,3}=  \frac{1}{32} \left(\Re\left[\zeta_{11}\right]-\Re\left[\zeta_{12}\right]+\Re\left[\zeta_{17}\right]-\Re\left[\zeta_{18}\right]\right), \qquad
  &&Z_ {1,1,2,2}=  \frac{1}{48} \left(\zeta_6-6 \Re\left[\zeta_{10}\right]\right),\\ \nonumber 
  &Z_ {1,1,2,3}=  \frac{1}{96} \left(-3 \Im\left[\zeta_{11}\right]+3 \Im\left[\zeta_{12}\right]-\Im\left[\zeta_{17}\right]+\Im\left[\zeta_{18}\right]\right), \qquad
  &&Z_ {1,1,3,3}=  \frac{1}{96} \left(2 \Re\left[\zeta_{14}\right]+2 \Re\left[\zeta_{15}\right]-2 \Re\left[\zeta_{16}\right]+\zeta_7+\zeta_8-\zeta_9\right), \\ \nonumber 
  &Z_ {1,2,2,2}=  \frac{1}{16} \left(2 \Im\left[\zeta_{10}\right]-\Im\left[\zeta_{13}\right]\right), \qquad
  &&Z_ {1,2,2,3}=  \frac{1}{96} \left(-3 \Re\left[\zeta_{11}\right]+3\Re\left[\zeta_{12}\right]+\Re\left[\zeta_{17}\right]-\Re\left[\zeta_{18}\right]\right), \\ \nonumber 
  &Z_ {1,2,3,3}=  \frac{1}{48} \left(-\Im\left[\zeta_{14}\right]-\Im\left[\zeta_{15}\right]+\Im\left[\zeta_{16}\right]\right), \qquad
  &&Z_ {1,3,3,3}=  \frac{1}{32} \left(\Re\left[\zeta_{19}\right]-\Re\left[\zeta_{20}\right]-\Re\left[\zeta_{21}\right]+\Re\left[\zeta_{22}\right]\right), \\ \nonumber 
  &Z_ {2,2,2,2}=  \frac{1}{16} \left(2 \Re\left[\zeta_{10}\right]-2\Re\left[\zeta_{13}\right]+\zeta_6\right), \qquad
  &&Z_ {2,2,2,3}=  \frac{1}{32} \left(\Im\left[\zeta_{11}\right]-\Im\left[\zeta_{12}\right]-\Im\left[\zeta_{17}\right]+\Im\left[\zeta_{18}\right]\right), \\ \nonumber 
  &Z_{2,2,3,3}=  \frac{1}{96} \left(-2 \Re\left[\zeta_{14}\right]-2\Re\left[\zeta_{15}\right]+2 \Re\left[\zeta_{16}\right]+\zeta_7+\zeta_8-\zeta_9\right), \qquad
  &&Z_ {2,3,3,3}=  \frac{1}{32} \left(-\Im\left[\zeta_{19}\right]+\Im\left[\zeta_{20}\right]+\Im\left[\zeta_{21}\right]-\Im\left[\zeta_{22}\right]\right), \\ 
  &Z_ {3,3,3,3}=  \frac{1}{16} \left(\zeta_1+\zeta_2+\zeta_3-\zeta_4-\zeta_5\right).
  \label{Zelements}
\end{flalign*}}
All other elements are zero, as a consequence of the $\mathrm{U(1)}_Y$ symmetry.

\section{Symmetric 2HDMEFT Potentials from Bilinear Formalism}\label{ap:coadjoint}

The maximal symmetry group in the $\mathbf{\Phi}$-space is $\mathrm{Sp(4)}$ and is generated by 
the 10 Lie generators${}$~\cite{apilaftsis}
\begin{align}
&K^{0,1,3}=\dfrac{1}{2}\sigma^3 \otimes \sigma^{0,1,3} \otimes \sigma^0, \, \quad \quad \;\;\;
K^2=\dfrac{1}{2}\sigma^0 \otimes \sigma^2 \otimes \sigma^0,\nonumber \\
&K^{4,5,8}=\dfrac{1}{2}\sigma^1 \otimes \sigma^{0,3,1} \otimes \sigma^0, \quad \quad 
K^{6,7,9}=\dfrac{1}{2}\sigma^2 \otimes \sigma^{0,3,1} \otimes \sigma^0\,.
\end{align}
The above generators possess the following Lie commutation relations:
\begin{equation}
    \big[K^a,\Sigma^A\big]=2if^{aAB}\Sigma^B, \qquad \big[K^a,K^b\big]=2ig^{abc}K^c, \qquad \big[\Sigma^A,\Sigma^B\big]=2ih^{ABa}K^a.
\end{equation}

Let us consider a transformation of the 6-vector $R^A$ under $\mathrm{Sp(4)}$. Since $R^0$ is invariant under Sp(4), we only consider variations in the `spatial' components of $R^A$, assuming that $A = 1,2,3,4,5$ in what follows unless otherwise stated. In the infinitesimal limit, we may evaluate
the variations 
 \begin{equation}
    \delta R^A\: =\: i\theta^a\big[K^a,R^A\big]\: =\: i\theta^a\mathbf{\Phi}^{\dagger}\big[K^a,\Sigma^A\big]\mathbf{\Phi}\: =\: 2\theta^af^{aAB}R^B\,. 
 \end{equation}
Thus, we find that in the reduced 5-dimensional bilinear $R^A$-space, 
the generators of the maximal symmetry group are the bi-adjoint matrices given by
 \begin{equation}
    \label{eq:Bi-adjoint}
    (T^a)_{AB}\: =\: -if^{aAB}\: =\: \mathrm{Tr}\big(\big[\Sigma^A,K^a\big]\Sigma^B\big)\,.
\end{equation}

With the help of  relation~\eqref{eq:Bi-adjoint}, we may derive the following 10
generators in the bi-adjoint representation of Sp(4):
 \begin{align}
     \label{eq:Ta}
T^0 &=\scriptsize \begin{pmatrix}
0 & 0 & 0 & 0 & 0 \\
0 & 0 & 0 & 0 & 0 \\
0 & 0 & 0 & 0 & 0 \\
0 & 0 & 0 & 0 & i \\
0 & 0 & 0 & -i & 0 
\end{pmatrix}, \qquad 
T^1 =\begin{pmatrix}
0 & 0 & 0 & 0 & 0 \\
0 & 0 & -i & 0 & 0 \\
0 & i & 0 & 0 & 0 \\
0 & 0 & 0 & 0 & 0 \\
0 & 0 & 0 & 0 & 0 \\
\end{pmatrix}, \qquad 
T^2=\begin{pmatrix}
0 & 0 & i & 0 & 0 \\
0 & 0 & 0 & 0 & 0 \\
-i & 0 & 0 & 0 & 0 \\
0 & 0 & 0 & 0 & 0 \\
0 & 0 & 0 & 0 & 0 
\end{pmatrix},\nonumber
\\[.2mm]
T^3 &=\scriptsize\begin{pmatrix}
0 & -i & 0 & 0 & 0 \\
i & 0 & 0 & 0 & 0 \\
0 & 0 & 0 & 0 & 0 \\
0 & 0 & 0 & 0 & 0 \\
0 & 0 & 0 & 0 & 0 
\end{pmatrix}, \qquad
T^4=\begin{pmatrix}
0 & 0 & 0 & 0 & 0 \\
0 & 0 & 0 & -i & 0 \\
0 & 0 & 0 & 0 & 0 \\
0 & i & 0 & 0 & 0 \\
0 & 0 & 0 & 0 & 0 
\end{pmatrix}, \qquad 
T^5=\begin{pmatrix}
0 & 0 & 0 & 0 & i \\
0 & 0 & 0 & 0 & 0 \\
0 & 0 & 0 & 0 & 0 \\
0 & 0 & 0 & 0 & 0 \\
-i & 0 & 0 & 0 & 0 
\end{pmatrix},\nonumber\\[.2mm]
T^6 &=\scriptsize\begin{pmatrix}
0 & 0 & 0 & 0 & 0 \\
0 & 0 & 0 & 0 & i \\
0 & 0 & 0 & 0 & 0 \\
0 & 0 & 0 & 0 & 0 \\
0 & -i & 0 & 0 & 0 
\end{pmatrix}, \qquad 
T^7 =\begin{pmatrix}
0 & 0 & 0 & i & 0 \\
0 & 0 & 0 & 0 & 0 \\
0 & 0 & 0 & 0 & 0 \\
-i & 0 & 0 & 0 & 0 \\
0 & 0 & 0 & 0 & 0 
\end{pmatrix}, \qquad
T^8 =\begin{pmatrix}
0 & 0 & 0 & 0 & 0 \\
0 & 0 & 0 & 0 & 0 \\
0 & 0 & 0 & 0 & -i \\
0 & 0 & 0 & 0 & 0 \\
0 & 0 & i & 0 & 0 
\end{pmatrix}, \nonumber\\[.2mm]
T^9 &=\scriptsize\begin{pmatrix}
0 & 0 & 0 & 0 & 0 \\
0 & 0 & 0 & 0 & 0 \\
0 & 0 & 0 & -i & 0 \\
0 & 0 & i & 0 & 0 \\
0 & 0 & 0 & 0 & 0
\end{pmatrix}. 
\end{align}
Note that these generators are identical to those of the SO(5) group
in the fundamental representation, thereby establishing the local
group isomorphism: Sp(4) $\cong$ SO(5).

The generators of the continuous symmetries in the $\mathbf{\Phi}$-space and their isomorphic partner in the $R^A$-space are shown in Table \ref{tab:gens}.

\addtocounter{table}{-1}
\begin{table}[t]
\begin{longtable}{ | s | s | s |}
\hline
\small{Symmetry in $\Phi$-space} & \small{Symmetry in $R^A$-space} & \small{Generators $T^a\leftrightarrow K^a$ and $D$} \\
\hline \hline
$\mathrm{U(1)_{PQ}}$ & $\mathrm{O(2)}\oplus\mathrm{O(2)} $ & $T^3$, $T^0$ \\ \hline
$\mathrm{CP1}\otimes\mathrm{SO(2)}$ & $\mathrm{CP1}\otimes\mathrm{O'(2)}\oplus\mathrm{O(2)} $ & $T^2$, $T^0$, $D_{\mathrm{CP1}}$ \\ \hline
$\mathrm{SU(2)_{HF}}$ & $\mathrm{O(3)}\oplus\mathrm{O(2)} $ & $T^{1,2,3}$, $T^0$ \\ \hline
$\mathrm{Sp(2)}_{\phi_1+\phi_2}$ & $\mathrm{SO(3)}$ & $T^{0,4,6}$ \\ \hline
$\mathrm{U(1)_{PQ}}\otimes\mathrm{Sp(2)}_{\phi_1\phi_2}$ & $\mathrm{O(2)}\oplus\mathrm{O(3)}$ & $T^3$, $T^{0,8,9}$ \\ \hline
$\mathrm{Sp(2)}_{\phi_1}\otimes\mathrm{Sp(2)}_{\phi_2}$ & $\mathrm{SO(4)}$ & $T^{0,3,4,5,6,7}$ \\ \hline
$\mathrm{Sp(4)}$ &   $\mathrm{SO(5)}$ & $T^{0,1,2,\dots,9}$ \\ \hline
\end{longtable}
\caption{\it Generators of the continuous accidental symmetries of the 2HDMEFT potential in $\Phi$/$R^A$-spaces.
Note that a  distinction between $\mathrm{O(2)}$ and $\mathrm{O(2)}'$ is made here, since these two-dimensional rotational groups act on the $R^{1,2}$ and $R^{1,3}$ spaces, respectively. }
\label{tab:gens}
\end{table}

In the {\em complete} 6-dimensional $R^A$-space (with $A=0,1,2,3,4,5$), under a given symmetry transformation in Sp(4), the tensors in the 2HDMEFT potential transform as \begin{align}
    \label{eq:transforms}
 &M_A\to M'_A = M_{A'} O^{A'}_{\;A}\,,  \qquad  &&L_{AB}\to L'_{AB} = L_{A'B'} O^{A'}_{\;A} O^{B'}_{\;B}\,,  \\
 &K_{ABC}\to K'_{ABC} = K_{A'B'C'} O^{A'}_{\;A} O^{B'}_{\;B} O^{C'}_{\;C}\,, \qquad &&Z_{ABCD}\to 
                                                                                       Z'_{ABCD} = Z_{A'B'C'D'}  O^{A'}_{\;A} O^{B'}_{\;B} O^{C'}_{\;C} O^{D'}_{\;D}\,.\nonumber
\end{align}
Note that $\{ O^{A'}_{\;A}\}$ is a $6\times 6$-dimensional matrix representation of $\mathrm{SO}(5) \subset \mathrm{SO}(1,5)$, such that $O^0_{\;A} = O^A_{\;0} = \delta^0_{\;A}$ in terms of the
6-dimensional Kronecker delta $\delta^A_{\;B}$, with $A,B=0,1,2,3,4,5$. 
In order that the effective potential is invariant under a given symmetry, we must demand that all tensors in~\eqref{eq:transforms} be invariant under this transformation,
 i.e.~$M'_A = M_A$, $L'_{AB} = L_{AB}$,  $K'_{ABC} = K_{ABC}$, $Z'_{ABCD} = Z_{ABCD}$ etc. 

 Alternatively, invariance under a given {\em infinitesimal} continuous symmetry also implies a set of relationships that involve the corresponding Lie-algebra generators $T^a$. Hence, for the effective potential to be invariant under a continuous symmetry group $G$, we must have \begin{align}
   &M_A \big[T^a\big]^{A'}_{\;A} =0, \quad 
L_{A'B} \big[T^a\big]^{A'}_{\;A}+L_{AB'} \big[T^a\big]^{B'}_{\;B} =0, \\
   &K_{A'BC} \big[T^a\big]^{A'}_{\;A} +K_{AB'C}  \big[T^a\big]^{B'}_{\;B} +K_{ABC'}  \big[T^a\big]^{C'}_{\;C} =0, \\
   &Z_{A'BCD}  \big[T^a\big]^{A'}_{\;A} +Z_{AB'CD} \big[T^a\big]^{B'}_{\;B} +Z_{ABC'D}\big[T^a\big]^{C'}_{\;C} +Z_{ABCD'}\big[T^a\big]^{D'}_{\;D} =0\,,
\end{align}
for all $T^a\in\mathfrak{g}$, where $\mathfrak{g}$ is the set of generators associated with a $6\times6$-dimensional representation of the Lie group~$G \subseteq \mathrm{SO}(5)$. Such a representation can be trivially obtained by adding an extra column and row of `0' in the matrices listed in~\eqref{eq:Ta}.

The above results can be generalised to include higher rank-$n$ symmetric tensors, $\Gamma^{(\mathrm{dim}=2n)}_{A_1,A_2\ldots, A_n}$, which will transform as
 \begin{equation}
   \Gamma^{(\mathrm{dim}=2n)}_{A_1,A_2\ldots, A_n}\to\  \Gamma'^{(\mathrm{dim}=2n)}_{A_1,A_2\ldots, A_n} =\: \Gamma^{(\mathrm{dim}=2n)}_{A_1',A_2'\ldots, A_n'} O^{A_1'}_{\;A_1} O^{A_2'}_{\;A_2} \ldots O^{A_n'}_{\;A_n}\,.
 \end{equation}
Thus, for $\Gamma^{(\mathrm{dim}=2n)}_{A_1,A_2\ldots, A_n}$ to be invariant under a continuous symmetry $G$, we require
 \begin{equation}
     \sum_{i=1}^{n} \Gamma^{(\mathrm{dim}=2n)}_{A_1,A_2\ldots, A_n} \big[ T^a \big]^{A_i}_{\;A_i'} =0\,,\quad \forall \;T^a\in\mathfrak{g}\, .
 \end{equation}
Hence, any symmetric 2HDMEFT potential can be calculated to any dimension $2n$ given the symmetry group generators.

\section{Discrete Transformation in Bilinear Formalism}\label{ap:dis} 
 
Having identified all the discrete symmetries in the 2HDMEFT framework in Section~\ref{sec:dis}, we now obtain the transformation matrices of discrete symmetries in the reduced bilinear $R^{A}$-space,
with $A=1,2,3,4,5$. In this spatially restricted bilinear field space, the usual 2HDM discrete symmetry transformations are given by 
\begin{align}
    \mathrm{D}_{Z_2}&=\mathrm{diag}(-1,-1,1,-1,-1)\,, \\
    \mathrm{D}_{\mathrm{CP1}}&=\mathrm{diag}(1,-1,1,1,-1)\,,\\
    \mathrm{D}_{\mathrm{CP2}}&=\mathrm{diag}(-1,-1,-1,1,-1)\,.
\end{align}

For a general discrete HF transformation, $Z_n$, the transformation matrix in the 
`spatial' part of the $R^A$-space can be represented as  
\begin{equation}
\mathrm{D}_{Z_n}=\begin{pmatrix}W_n&&\\&1&\\&&W_n\end{pmatrix}, \qquad W_n=\begin{pmatrix}\mathrm{Re}(\omega_n)&-\mathrm{Im}(\omega_n)\\\mathrm{Im}(\omega_n)&
\mathrm{Re}(\omega_n)\end{pmatrix},
\end{equation}
with $\omega_n=e^{2\pi i/n}$. Therefore, in this reduced $R^A$-space, the transformation matrices for the discrete accidental HF symmetries present in the 2HDMEFT potential extended up to dimension~8 can be written as
\begin{align}
    \mathrm{D}_{Z_3}=\begin{pmatrix}
    -\frac{1}{2}&-\frac{\sqrt{3}}{2}&0&0&0\\ 
    \frac{\sqrt{3}}{2}&-\frac{1}{2}&0&0&0\\
    0&0&1&0&0\\
    0&0&0&-\frac{1}{2}&-\frac{\sqrt{3}}{2}\\ 
    0&0&0&\frac{\sqrt{3}}{2}&-\frac{1}{2}
    \end{pmatrix}, \qquad\qquad 
\mathrm{D}_{Z_4}=\begin{pmatrix}
    0&-1&0&0&0\\ 
    1&0&0&0&0\\
    0&0&1&0&0\\
    0&0&0&0&-1\\ 
    0&0&0&1&0
    \end{pmatrix}.
\end{align}

By analogy, the general CP$n$ transformations can be constructed from the product
\begin{equation}
D_{\text{CP}n}= D_{Z_n} \cdot D_{S_2} \cdot  D_{\mathrm{CP1}}\,,
\end{equation}
where
\begin{equation} 
D_{S_2}=\mathrm{diag}(1,-1,-1,-1,-1).
\end{equation}
Hence, in the bilinear `spatial' $R^{A}$-space, the transformation matrices associated with CP3 and CP4 discrete symmetries are given by
\begin{align}
D_{\text{CP3}}={1\over 2}\begin{pmatrix}
-1&-\sqrt{3}  & 0& 0 & 0 \\
\sqrt{3}  &-1 & 0 & 0 &0\\
0&0 &-2 & 0  &0 \\
0&0 & 0 &1&-\sqrt{3} \\
0&0& 0&-\sqrt{3}  &-1\\
\end{pmatrix}, \qquad D_{\text{CP4}}&=\begin{pmatrix}
0&-1 &0 & 0 &0 \\
1  &0& 0 & 0 & 0\\
0 & 0 &-1 & 0 & 0 \\
0 &0& 0 &0&-1 \\
0&0& 0 &-1 &0\\
\end{pmatrix}.
\end{align}
The non-Abelian tetrahedral group, $D_n$, transformations are generated by the set of generators corresponding to $Z_n$ and $S_2$, and their product
\begin{equation}
    D_{Z_n}, \quad D_{S_2}, \quad D_{Z_n}D_{S_2}.
\end{equation}

We note that the parameter relations of the accidentally symmetric 2HDMEFT potentials can be obtained via analysis of the corresponding group generators, in a fashion similar to what was done in Appendix~\ref{ap:coadjoint}.
 
\vspace{1cm}

\bibliographystyle{unsrt}
\bibliography{2HDMEFT.bib}

\end{document}